\documentclass[twocolumn,pra,aps,showpacs,amsmath,amssymb,floatfix]{revtex4-2}

\usepackage{url}
\usepackage{amsmath}
\usepackage{amssymb}
\usepackage{graphicx}
\usepackage{hyperref}
\usepackage{dcolumn}
\usepackage{bm}
\usepackage{xcolor}
\usepackage{booktabs}
\usepackage{etoolbox}
\robustify\itshape
\usepackage{siunitx}
\usepackage[caption=false]{subfig}
\usepackage{multirow}
\hypersetup{
	colorlinks=true,       
	linkcolor=blue,        
	citecolor=blue,        
	filecolor=magenta,     
	urlcolor=blue         
}

\newcommand{\tdl}{\tau_\mathrm{DL}}
\newcommand{\tfl}{\tau_\mathrm{FL}}
\newcommand{\tdltm}{\tau^{(1)}_{2\bar2}}
\newcommand{\tfltm}{\tau^{(2)}_{2\bar2}}
\newcommand{\bdl}{B_{\mathrm{DL}}}
\newcommand{\bfl}{B_{\mathrm{FL}}}
\newcommand{\bdltm}{B_{3m\mathrm{DL}}}
\newcommand{\bfltm}{B_{3m\mathrm{FL}}}

\begin{document}

\title{Spin-orbit torque in a three-fold-symmetric bilayer and its effect on magnetization dynamics}
\author{Wuzhang Fang}
\thanks{These two authors contributed equally.}
\author{Edward Schwartz}
\thanks{These two authors contributed equally.}
\author{Alexey A. Kovalev}

\author{K. D. Belashchenko}
\affiliation{Department of Physics and Astronomy and Nebraska Center for Materials and Nanoscience, University of Nebraska-Lincoln, Lincoln, Nebraska 68588, USA}

\begin{abstract}
    Field-free switching of perpendicular magnetization has been observed in an epitaxial L1$_1$-ordered CoPt/CuPt bilayer and attributed to spin-orbit torque (SOT) arising from the crystallographic $3m$ point group of the interface. Using a first-principles nonequilibrium Green’s function formalism combined with the Anderson disorder model, we calculate the angular dependence of the SOT in a CoPt/CuPt bilayer and find that the magnitude of the $3m$ SOT is about 20\% of the conventional dampinglike SOT. We further study the magnetization dynamics in perpendicularly magnetized films in the presence of $3m$ SOT and Dzyaloshinskii-Moriya interaction, using the equations of motion for domain wall dynamics and micromagnetic simulations. We find that for systems with strong interfacial DMI characterized by the N\'eel character of domain walls, a very large current density is required to achieve deterministic switching because reorientation of the magnetization inside the domain wall is necessary to induce the switching asymmetry. For thicker films with relatively weak interfacial DMI and the Bloch character of domain walls the deterministic switching with much smaller currents is possible, which agrees with recent experimental findings. 
\end{abstract}

\maketitle
\section{Introduction}
Spin-orbit torque (SOT) \cite{Manchon-RMP} provides an efficient way of controlling the magnetization in spintronic devices, such as the magnetic random access memories. In ferromagnet (FM)/heavy-metal (HM) bilayers, the spin Hall effect in a non-magnetic layer with strong spin-orbit interaction (SOI) can be used to produce a flow of angular momentum into neighbouring ferromagnet, and, thus, induce SOT. 
Various mechanisms can contribute to SOT and many experiments are consistent with the spin-Hall effect~\cite{RevModPhys.87.1213,PhysRevLett.106.036601,Liu2012,PhysRevLett.109.096602,Zhu2021} mechanism of the dampinglike SOT, and the inverse spin-galvanic effect mechanism~\cite{Chernyshov2009,MihaiMiron2010,Miron2011,Manchon2015} of the fieldlike SOT. Other mechanisms of SOT have also been identified such as 
the orbital Hall effect~\cite{PhysRevResearch.2.033401,PhysRevResearch.2.013177}, the planar Hall effect~\cite{Safranski2018,PhysRevLett.124.197204}, the magnetic spin-Hall effect~\cite{PhysRevB.99.220405,PhysRevResearch.2.023065}, and various interfacial mechanisms, e.g., associated with interfacial spin current generation~\cite{PhysRevMaterials.3.011401,Amin2020,Kirill2}.

Current-induced magnetization switching by SOT has been demonstrated in bilayers with in-plane magnetization~\cite{Liu2012}. Current-induced switching of perpendicular magnetization in systems with axial symmetry requires additional symmetry breaking mechanism, such as application of external magnetic field~\cite{Miron2011,PhysRevLett.109.096602,Liu2012}. Alternatively, to enable the field-free switching of perpendicular magnetization one can resort to systems of trilayers~\cite{Humphries2017,Baek2018}, or systems with lower symmetry~\cite{Yu2014,MacNeill2016,Liu2021,Zhu2023}. A possibility of field free switching of perpendicular magnetization in a system with lower crystal symmetry has been demonstrated in a recent experiment~\cite{Liu2021}.

In this work, we consider a bilayer system with lower crystal symmetry, which, as has been demonstrated in Ref.~\cite{PhysRevB.102.014401}, can display spin Hall effect with unconventional polarization, and unconventional SOT. Developing these ideas further, one can expand SOT in orthogonal vector spherical harmonics, identifying only those contributions that are allowed by symmetry~\cite{Kirill2,Xue2023}. The CoPt/CuPt bilayer considered here has C$_{3v}$ symmetry, which allows the existence of so-called $3m$ dampinglike and fieldlike torques \cite{Liu2021,Ovalle}.

We begin by performing first-principles calculations of the spin-orbit torquances in an L1$_1$-ordered CoPt/CuPt bilayer and find appreciable $3m$ field-like and damping-like torques in addition to standard damping-like and field-like torques. We find that the $3m$ damping-like torque decreases significantly at higher disorder strengths, while the $3m$ field-like torque is largely unaffected by disorder.

We then examine the magnetization reversal process in a disordered magnetic system, as a result of these torques. We use a collective coordinate model to demonstrate that the $3m$ damping-like torque results in domain wall expansion in a way analogous to an out-of-plane applied field, while the effect of the $3m$ field-like torque is more complicated. For N\'eel-type domain walls, the $3m$ field-like torque produces no net expansion of an isolated domain. In contrast, Bloch walls experience a finite average pressure which is approximately quadratic in the applied current density for arbitrarily small $3m$ torque. Our collective coordinate model is supported by micromagnetic simulations.

\section{First-principles calculations}
\label{sec:fp}
\subsection{Computational details}
\label{sec:computational}

\begin{figure}
    \includegraphics[width=0.9\columnwidth]{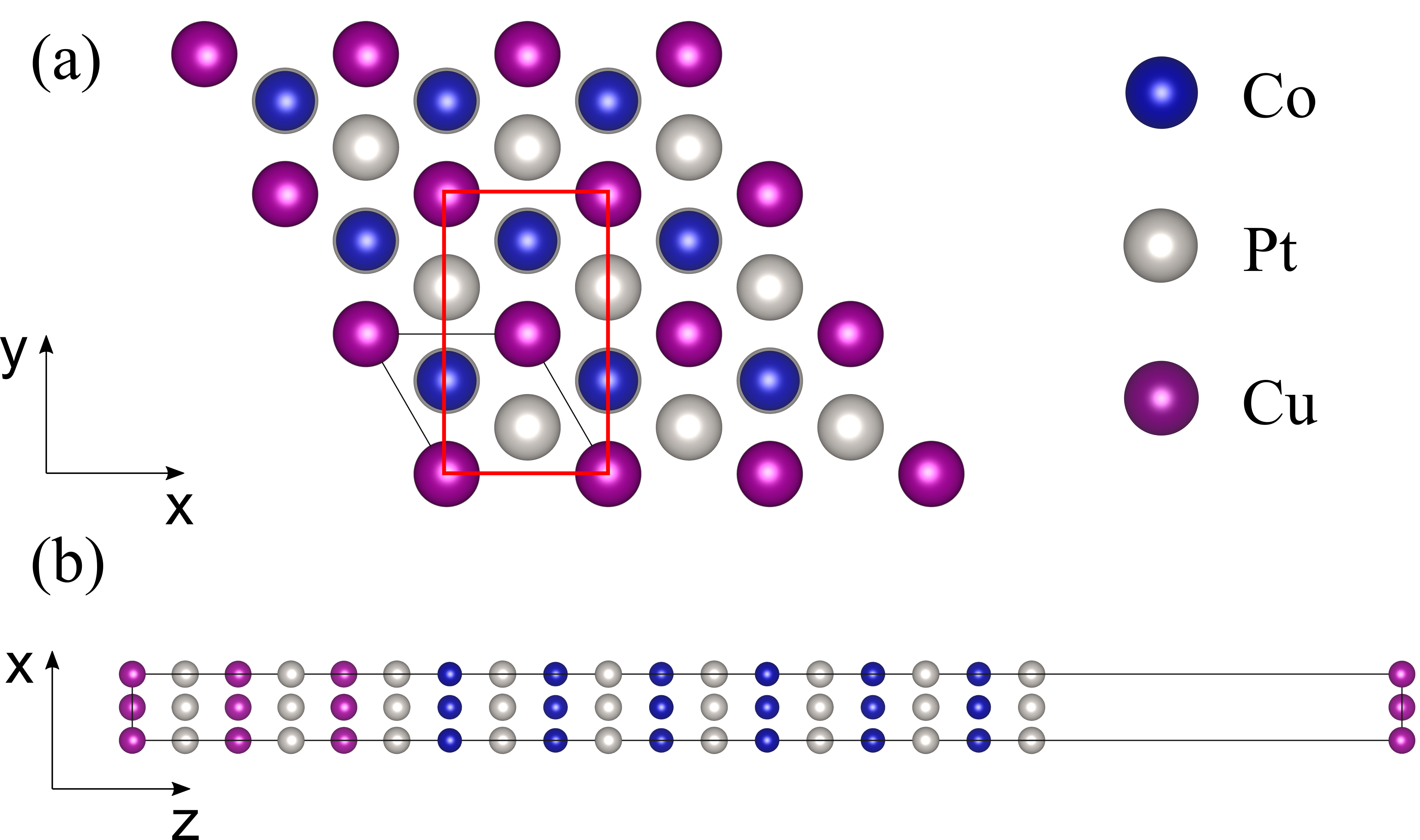} 
\caption{(a) Top and (b) side view of an L1$_1$-ordered CoPt/CuPt bilayer. Red rectangle: unit cell used in the transport calculations.}
\label{fig:structure}
\end{figure}

We consider an L1$_1$-ordered CoPt/CuPt bilayer for which the torques with higher-order symmetry were studied experimentally \cite{Liu2021}. We take this bilayer to include six monolayers of CuPt and twelve of CoPt, as shown in Fig. \ref{fig:structure}. This system has $C_{3v}$ (or $3m$ in Hermann–Mauguin notation) symmetry. We performed a structural optimization of a smaller CoPt(6)/CuPt(6) bilayer using the projector-augmented wave (PAW) method \cite{BLOCHL1994} implemented in the Vienna Ab Initio Simulation Package (VASP) \cite{VASP1,VASP2,VASP3} with the Perdew-Burke-Ernzerhof (PBE) \cite{PBE} exchange-correlation functional. The resulting in-plane lattice constant is \SI{2.702}{\angstrom}, and the interlayer spacings vary from 2.12 to 2.19 \AA. For simplicity, we fixed the interlayer spacings at 2.15 \AA\ in the transport calculations. The magnetic moments of Co atoms in the CoPt layer and at the free surface are 1.86 and 1.89 $\mu_B$, respectively. The magnetic moments of Pt atoms in the CoPt layer and at the interface are 0.26 and 0.15 $\mu_B$, respectively. The magnetic moments of Cu and Pt atoms in the CuPt layer are negligible. 

Transport calculations were performed using the nonequilibrium Green's function (NEGF) technique implemented within the atomic sphere approximation of the tight-binding linear muffin-tin orbital (TB-LMTO) method \cite{AndersenLMTO,Turek} in the Questaal code \cite{Faleev2005,QUESTAAL,Kirill1,Kirill2}. Both Fermi-surface and Fermi-sea contributions were considered \cite{Kirill1}. 

The scattering region was taken to be a rectangular cuboid supercell obtained by periodically repeating the unit cell shown by the red rectangle in Fig.\ \ref{fig:structure}. The boundary conditions in the $yz$ plane are periodic, and the current flows in the $\hat{x}$ direction, which is perpendicular to a mirror plane. The vacuum region is represented by six monolayers of empty spheres. The size of the supercell is denoted as $L_x\times L_y$, where $L_x$ and $L_y$ are the numbers of unit cells included in the supercell along the $\hat{x}$ and $\hat{y}$ directions, respectively.

Disorder was simulated within the Anderson model as a uniformly distributed random potential $V_i$, $-V_m<V_i<V_m$, applied on each atomic site $i$. To account for the potential drop in the contacts, the electric field in the embedded region is found as $E=V/L_\mathrm{eff}$, where $V$ is the voltage drop between the left and right leads, $L_\mathrm{eff}=R(dR/dL)^{-1}$, and $R=1/G_{LB}$ is the Landauer-B\"uttiker resistance of the active region of thickness $L$ \cite{CoPt_2021}.

For the Fermi-surface contribution, we considered three disorder strengths: $V_m=0.65$, 0.82, and 0.98 eV, which yield resistivities of 13, 19, and \SI{26}{\micro\ohm\centi\meter}, respectively. Using supercells with $L_y=1$, 2, and 4, we found that $L_y=2$ (which corresponds to 12 monolayers, as seen in Fig. \ref{fig:structure} is sufficient to converge the $3m$ torques, which are of central interest here. Unless otherwise noted, torquances were calculated using $150\times 2$ supercells. Averaging was performed over 150 disorder configurations at $V_m=0.65$ eV and over 36 configurations at $0.82$ and 0.98 eV.

To estimate the Fermi-sea contribution to the DL and $3m$ torques, we used a $60\times 1$ supercell with 100 disorder configurations at $V_m=0.65$ eV. A finite bias of order 0.1 mV was applied symmetrically, and the torques at a positive and negative bias were subtracted to remove the equilibrium torque associated with disorder-dependent magnetic anisotropy. The Fermi-sea contribution involves an integral over the filled states, which was evaluated on an elliptical contour shifted vertically by 1 mRy along the imaginary axis. A 101-point Legendre quadrature was used on the semi-ellipse, and a uniform 10-point mesh on each vertical segment.

The total torquances acting on the magnetization of the whole bilayer will be expressed per unit area of the interface, in units of \SI{e5}{} $(\hbar/2e)$ \SI{}{\per\ohm\per\meter}. For the DL and FL torques this definition is identical to the torque efficiencies $\xi^E_\mathrm{DL}$, $\xi^E_\mathrm{FL}$ that are often used in experimental reports.

The intrinsic spin Hall conductivities of bulk CoPt and CuPt were calculated using Wannier interpolation \cite{Wannier90,Wannier90-SHC} combined with VASP calculations using the same crystal structure as in our bilayer model. We consider the tensor component $\sigma_{zx}^{y}$, which corresponds to spin current with polarization $\hat{y}$ flowing in the $\hat{z}$ direction under an electric field applied in the $\hat{x}$ direction.

\subsection{Angular dependence}
\label{sec:angular}

The dependence of the (total or site-resolved) torque $\mathbf{T}(\mathbf{m})$ on the orientation of the magnetization unit vector $\mathbf{m}$ can be conveniently represented using real vector spherical harmonics (VSH) $\mathbf{Z}_{lm}^{(\nu)}(\mathbf{m})$, $\nu\in\{1,2\}$ \cite{Kirill2}. Real VSH with $m<0$ are proportional to the imaginary part of the corresponding complex VSH, and those with $m>0$ to its real part. We will use a bar over a digit to denote negative values of $m$. All VSH with $\nu=1$ are purely dampinglike and those with $\nu=2$ purely fieldlike; the terms with even (odd) $\nu+l$ are even (odd) under time reversal \cite{Kirill2}.

VSH form a complete orthonormal basis set, and the lowest-order $\mathbf{Z}_{1\bar1}^{(1)}$ and $\mathbf{Z}_{1\bar1}^{(2)}$ harmonics correspond to the conventional DL and FL torques up to a normalization factor: 
$\mathbf{Z}_{1,\bar1}^{(1)}=a\mathbf{m}\times(\mathbf{y}\times\mathbf{m})$ and $\mathbf{Z}_{1,\bar1}^{(2)}=a\mathbf{y}\times\mathbf{m}$, where $a=\sqrt{3/(8\pi)}$. For easy matching with the DL and FL torquances, we will use a rescaled basis of $\tilde{\mathbf{Z}}_{lm}^{(\nu)}=a^{-1}\mathbf{Z}_{lm}^{(\nu)}$ in the following. Thus, the (total or site-resolved) torquance $\boldsymbol{\tau}(\mathbf{m})=\mathbf{T}(\mathbf{m})/E$ will be represented as
\begin{align}
\boldsymbol{\tau}(\mathbf{m})=\sum_{lm\nu}\tau^{(\nu)}_{lm}\tilde{\mathbf{Z}}_{lm}^{(\nu)}
\label{expansion}
\end{align}
and we will also use $\tdl\equiv\tau^{(1)}_{1\bar1}$ and $\tfl\equiv\tau^{(2)}_{1\bar1}$ to denote the torquances with $l=1$.

For a bilayer with axial $C_{\infty v}$ symmetry, only terms with $m=(-1)^{l}$ are allowed in the expansion (\ref{expansion}) \cite{Kirill2}. The $3m$ ($C_{3v}$) symmetry of the CoPt/CuPt system allows additional terms, as shown in Appendix \ref{app:symmetry}. With the current flowing in the $x$ direction, perpendicular to one of the mirror planes, these terms start with $\tau_{2,\bar2}^{(\nu)}$. According to the general rule \cite{Kirill2}, the torque represented by $\tilde{\mathbf{Z}}_{2,\bar2}^{(1)}$ is purely dampinglike and $\tilde{\mathbf{Z}}_{2,\bar2}^{(2)}$ purely fieldlike. The latter is identical with the ``$3m$ torque'' defined in Ref.\ \onlinecite{Liu2021}. We will refer to $\tilde{\mathbf{Z}}_{2,\bar2}^{(1)}$ and $\tilde{\mathbf{Z}}_{2,\bar2}^{(2)}$ as $3m$ DL and $3m$ FL torque, respectively. For the CoPt/CuPt bilayer studied in this paper, our first-principles calculations reveal that beyond the conventional DL and FL torquances, only $\tau_{2,\bar2}^{(1)}$, $\tau_{2,\bar2}^{(2)}$, and $\tau_{2,1}^{(2)}$ are appreciable.

The Fermi-surface contribution to the torquances was calculated for 32 orientations of $\mathbf{m}$ and projected on the VSH basis \cite{Kirill2}. The Fermi-sea contribution is only allowed  for time-reversal-even terms $\tau_\mathrm{DL}$,  $\tau_{2,\bar2}^{(2)}$, and $\tau_{2,1}^{(2)}$. We calculate this contribution only at $\mathbf{m}=\hat z$ and $\mathbf{m}=\hat y$. Although the torque at $\mathbf{m}=\hat z$ can have contributions from $\tau_\mathrm{DL}$ and $\tau^{(2)}_{21}$, we allocate it entirely to $\tau_\mathrm{DL}$, given that $\tau^{(2)}_{21}$ is expected to be relatively small. The $z$-component of the torque at $\mathbf{m}=\hat y$ is fully attributable to the $3m$ FL torque $\tau^{(2)}_{2\bar2}$.

\subsection{Exchange and spin-orbital torques}
\label{sec:xcso}

In the atomic sphere approximation, the Kohn-Sham Hamiltonian can be written as
\begin{equation}
H=-\frac{\nabla^2}{2} + \sum_i \left[V_i(\mathbf{r})+\boldsymbol{\sigma}\cdot\mathbf{B}^i_{xc}(\mathbf{r})+H^i_{SO}\right]
\end{equation}
where $V_i(\mathbf{r})$ and $\mathbf{B}^i_{xc}(\mathbf{r})$ describe, respectively, the spin-independent and spin-dependent parts of the spherically symmetric Kohn-Sham potential inside the atomic sphere at site $i$, and $H^i_{SO}$ is the spin-orbit coupling operator for site $i$.

The rate of change of the spin angular momentum $\mathbf{S}_i$ on atom $i$, which is zero in the steady state, is proportional to the expectation value of the commutator $[\mathbf{S}_i,H]$. The commutators with the four terms in $H$ give, respectively, the discrete divergence of spin current $\nabla_i \hat Q_s$, zero, the exchange torque $\overline{\mathbf{T}}^i_{xc}$, and the spin-orbital torque $\mathbf{T}^i_{SO}$. The bar in $\overline{{\mathbf{T}}}^i_{xc}$ indicates that this torque acts on the conduction electrons and is opposite in sign to the reaction torque ${\mathbf{T}}^i_{xc}$ that acts on the localized spin. The spin current tensor $\hat Q_s$ is well-defined (up to adding a curl) because the kinetic energy operator conserves spin angular momentum. 
In the steady state we have \cite{HaneyStiles2010,Go2020}
\begin{equation}
    \overline{\mathbf{T}}^i_{xc} + \mathbf{T}^i_{SO} + \nabla_i \hat Q_s = 0
    \label{continuity}
\end{equation}
The sum $\sum_i\nabla_i \hat Q_s$ taken over all atoms in a volume is equal to the total spin current flowing into that volume.
Consider a volume $\Omega$ between two planes that are both parallel to the embedding planes and far enough from them so that the disorder-averaged spin current passing through them is the same thanks to translational symmetry. Then we find from (\ref{continuity}) that $\sum_{i\in\Omega}\mathbf{T}^i_{xc}=\sum_{i\in\Omega}\mathbf{T}^i_{SO}$, i.e., the total exchange torque on the localized spins in volume $\Omega$ is equal to the total spin-orbital torque on the conduction electrons in that volume. Following Ref. \onlinecite{Go2020}, we will refer to these torques below as exchange torque and spin-orbital torque.

We calculate both $\mathbf{T}^i_{xc}$ and $\mathbf{T}^i_{SO}$ independently from the on-site blocks of the nonequilibrium spin density matrix. The ``sum rule'' for the exchange and spin-orbital torques helps check the convergence of the results with respect to the size of the embedded supercell, which may be delicate for the relatively small $3m$ torques. The sum rule only applies to the gauge-invariant total torque and not individually to the Fermi-surface and Fermi-sea contributions. On the other hand, $\mathbf{T}^i_{xc}$ and $\mathbf{T}^i_{SO}$ provide complementary site-resolved information: $\mathbf{T}^i_{xc}$ indicates where the magnetization torque is acting, and $\mathbf{T}^i_{SO}$ indicates where the angular momentum is exchanged between the spin and orbital degrees of freedom.

\subsection{First-principles results}
\label{sec:fpresults}

\begin{figure}[htb]
    \includegraphics[width=0.9\columnwidth]{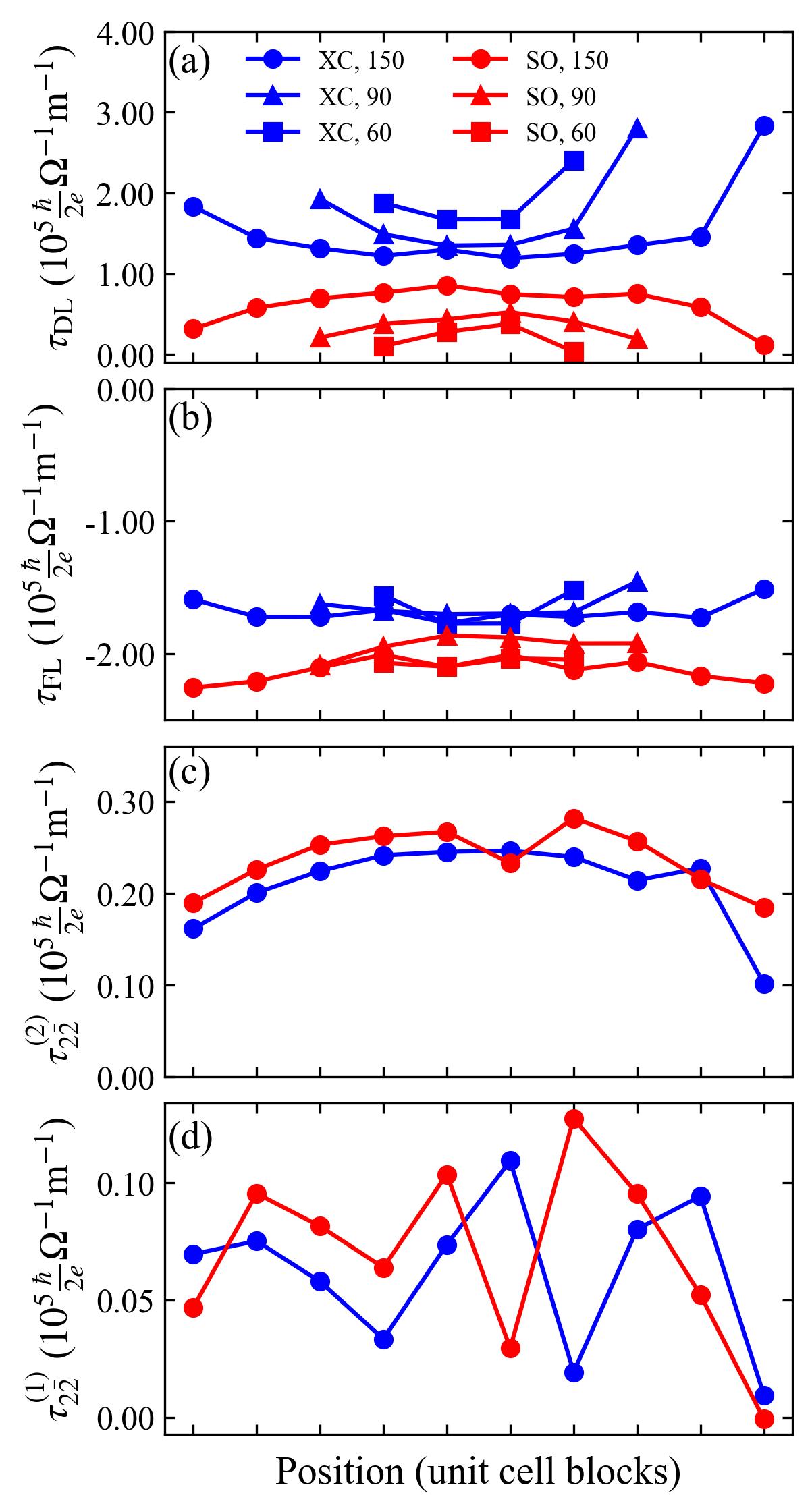}
\caption{(a) Dampinglike $\tdl$, (b) Fieldlike $\tfl$, (c) $3m$ fieldlike $\tfltm$, and (d) $3m$ dampinglike $\tdltm$ torquances in a CoPt(12)/CuPt(6) bilayer at $V_m=0.65$ eV as a function of the position along the current flow direction. Blue color (XC): exchange, red color (SO): spin-orbital torquances. Only the Fermi surface contribution is shown. Each point shows an average over a block of 15 unit cells.}
\label{fig:edge-effect}
\end{figure}

First we examine the convergence of the torquances with respect to the length of the scattering region.
Fig. \ref{fig:edge-effect} shows the Fermi-surface contribution for the exchange and spin-orbital torquances as a function of the position along the direction of the current flow at $V_m=0.65$ eV. In this plot, the supercell is divided lengthwise into equal 15-unit-cell blocks, and each data point shows the average torquance for one such block. The four panels correspond to 
$\tau_\mathrm{DL}$, $\tau_\mathrm{FL}$, $\tfltm$, amd $\tdltm$. 

Figure \ref{fig:edge-effect}(a), which also includes data for shorter $60\times2$ and $90\times2$ supercells, shows significant edge effects, with the DL torquance near the embedding planes being significantly different from the middle of the scattering region. The exchange torquance $\tdl^{xc}$ in the middle of the scattering region is very similar for $L_x=60$, 90, and 150, but the spin-orbital torquance $\tdl^\mathrm{SO}$ converges more slowly. Thus, we expect the exchange torquances extracted from these calculations to be more accurate. The edge effects should become less pronounced with increasing disorder due to the reduction of the mean-free path and the spin-diffusion length. When performing disorder averaging, we exclude 30 unit cells near each edge. Figures \ref{fig:edge-effect}(c) and \ref{fig:edge-effect}(d) suggest this should also be sufficient for the $3m$ fieldlike and dampinglike torques. Figure \ref{fig:edge-effect}(d) shows strong disorder sampling noise in the spatially resolved plot of the relatively small $3m$ DL torque, but it is consistently positive and not subject to strong edge effects. 

Next we examine the validity of the sum rule mentioned in Section \ref{sec:xcso}, according to which the total exchange and spin-orbital torques must be the same. Table \ref{tab:150x2} shows the calculated Fermi-surface contribution to the exchange and spin-orbital torquances for different disorder strengths, as well as the Fermi-sea contribution for $V_m=0.65$ eV. The latter only contributes to even torquances $\tdl$ and $\tfltm$.
We see that the Fermi sea contributions to $\tdl^{xc}$ and $\tdl^\mathrm{SO}$ are relatively small, but their inclusion improves the agreement between the total $\tdl^{xc}$ and $\tdl^\mathrm{SO}$ at $V_m=0.65$ eV. The remaining discrepancy of \SI{0.2e5}{} $(\hbar/2e)$ \SI{}{\per\ohm\per\meter} may be due to incomplete convergence of $\tdl^\mathrm{SO}$ with respect to the supercell size, which appears likely in view of the convergence analysis in Fig. \ref{fig:edge-effect}(a), and disorder sampling errors. Because the Fermi sea terms are rather small, we make an assumption that they depend weakly on disorder strength and use the values calculated at $V_m=0.65$ eV for all disorder strengths. We also see good agreement between the exchange and spin-orbital torquances for the FL, $3m$ FL, and $3m$ DL harmonics. In particular, for the $3m$ components they agree within the sampling noise.

The intrinsic spin Hall conductivity (SHC) of 
CuPt, calculated as described in Section \ref{sec:computational}, 
is \SI{1.44e5}{} $(\hbar/2e)$ \SI{}{\per\ohm\per\meter}, which is a factor of 3 smaller compared to Pt \cite{SHC-Pt}. Due to the suppression by backflow and the small thickness of CuPt, the resulting spin-Hall generated dampinglike SOT in CoPt would be expected to be significantly smaller, perhaps by a factor of 2--3. This rough estimate appears to be close to the calculated dampinglike SOT at $V_m=0.98$ eV, and a few times smaller than that for $V_m=0.65$ eV in CoPt/CuPt (see Table \ref{tab:150x2}). Thus, we conclude that $\tdl$ must have a large extrinsic contribution at smaller values of $V_m$.
\begin{table*}[htb]
\begin{tabular}{p{2cm}|c|l|c|S[table-format=3.2]S[table-format=3.3]ccc}
\hline
Contribution & Type & SOC & {$V_m$ (eV)} & {$\tdl$} & {$\tfl$} & {$\tdltm$} & {$\tfltm$} & {$\tau^{(2)}_{21}$}\\
\hline          
\multirow{6}{*}{Fermi surface} & \multirow{3}{*}{XC}& \multirow{6}{*}{Full}    & 0.65  & 1.26 & -1.71 &  0.06           & 0.23 & 0.27 \\  
                                &                     &                        & 0.82  & 0.85 & -1.79 & \textit{-0.01}  & 0.11 & 0.11 \\
                                &                     &                        & 0.98  & 0.63 & -1.69 & \textit{-0.05}  & 0.09 & \textit{0.01} \\
                                 \cline{4-9}
                                & \multirow{3}{*}{SO} &                        & 0.65  & 0.75 & -2.08 &  0.08           & 0.26 & 0.33 \\     
                                &                     &                        & 0.82  & 0.55 & -2.01 &  \textit{0.01}  & 0.12 & 0.17 \\
                                &                     &                        & 0.98  & 0.51 & -1.81 &  \textit{0.01}  & 0.10 & \textit{0.05}\\
\hline
\multirow{2}{*}{Fermi sea} & XC & \multirow{2}{*}{Full}& \multirow{2}{*}{0.65} & -0.22 & {-} & {-} &  0.00 & {-}\\   
                           & SO &                      &                       &  0.10 & {-} & {-} & -0.02 & {-}\\
\hline
\multirow{6}{*}{Total}          & \multirow{3}{*}{XC} &  \multirow{6}{*}{Full} & 0.65  & 1.04 & -1.71 &  0.06           & 0.23 & 0.27\\   
                                &                     &                        & 0.82  & 0.63 & -1.79 & \textit{-0.01}  & 0.11 & 0.11\\
                                &                     &                        & 0.98  & 0.41 & -1.69 & \textit{-0.05}  & 0.09 &\textit{0.01}\\
                                \cline{4-9}
                                & \multirow{3}{*}{SO} &                        & 0.65  & 0.85 & -2.08 &  0.08           & 0.24 & 0.33\\      
                                &                     &                        & 0.82  & 0.65 & -2.01 &  \textit{0.01}  & 0.10 & 0.17\\
                                &                     &                        & 0.98  & 0.61 & -1.81 &  \textit{0.01}  & 0.08 & \textit{0.05}\\
\hline
\multirow{3}{*}{\parbox{\linewidth}{Fermi surface\\$90\times2$}}& \multirow{3}{*}{XC} &   Full  & \multirow{3}{*}{0.65} & 1.43 & -1.67 &  0.11          &  0.24          & 0.33\\  
                                &                                                     &   CoPt  &                       & 0.41 &  0.49 &  0.08          &  0.20          & 0.30\\
                                &                                                     &   CuPt  &                       & 0.88 & -2.46 & \textit{-0.01} & \textit{-0.02} & \textit{0.01}\\
\hline                             
\end{tabular}
    \caption{Exchange (XC) and spin-orbital (SO) torquances in a CoPt(12)/CuPt(6) bilayer in units of \SI{e5}{} $(\hbar/2e)$ \SI{}{\per\ohm\per\meter}. The typical error bar from disorder sampling is \SI{0.04e5}{} in these units. Values in italics are not statistically significant.}
    \label{tab:150x2}
\end{table*}

\begin{figure}
    \includegraphics[width=0.85\columnwidth]{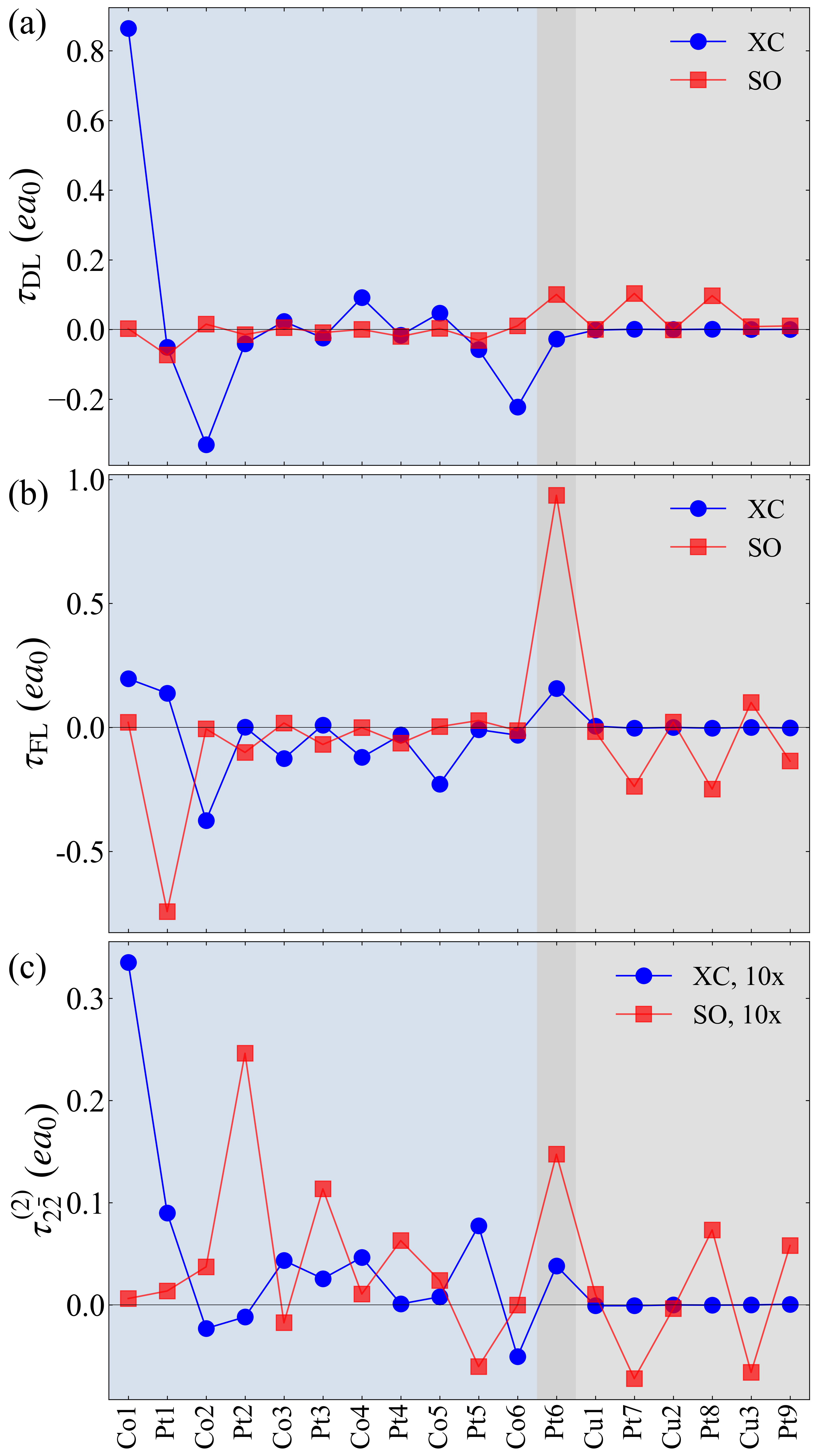} 
\caption{Site-resolved (a) dampinglike $\tau_{\mathrm{DL}}$, (b) fieldlike $\tau_{\mathrm{FL}}$, and (c) $3m$ fieldlike $\tfltm$ in a CoPt(12)/CuPt(6) bilayer at $V_m=0.65$ eV. Blue color (XC): exchange, red color (SO): spin-orbital torquances. Blue, dark grey, and light grey regions denote CoPt, the interfacial Pt layer, and CuPt, respectively.}
\label{fig:siteresolved}
\end{figure}

The fieldlike torquance $\tfl$ listed in Table \ref{tab:150x2} is nearly constant in the range of $V_m$ from 0.64 to 0.98 eV, which is unusual, given that it is expected to be dominated by the inverse spin-galvanic mechanism which scales with the conductivity. An explanation for this behavior is provided by the examination of site-resolved torquances. Figure \ref{fig:siteresolved}(b) reveals strong sources of the fieldlike torque, indicated by $\tfl^\mathrm{SO}$, near the free surface of CoPt and at the CoPt/CuPt interface, which have opposite signs and nearly cancel each other. As seen in Fig. \ref{fig:FL-site}(d) (see Appendix \ref{app:siteres}), each of these sources is strongly reduced by increasing disorder, as expected for the inverse spin-galvanic mechanism. Competition between these sources results in non-monotonic disorder dependence of the total fieldlike torquance.

The $3m$ FL torquance $\tfltm$ decreases with increasing disorder strength roughly proportional to the DL torquance, such that $\tfltm/\tdl\sim0.2$, using the calculated exchange torques. This observation suggests that $\tfltm$ also includes a substantial extrinsic contribution.

To get further insight into the mechanism of SOT, we recalculated the Fermi-surface contributions to the torquances with spin-orbit coupling turned on only in CuPt or only in CoPt (including the interfacial Pt atom), at $V_m=0.65$ eV. These calculations used a smaller $90\times 2$ supercell. The results, shown at the bottom of Table \ref{tab:150x2}, show that all torques generated by spin-orbit coupling in CoPt and CuPt are approximately additive. Furthermore, we note that $\tfltm$ becomes essentially zero if spin-orbit coupling is turned off in CoPt, including the interfacial Pt atom. The plot of site-resolved spin-orbital $3m$ FL torquances in Fig.\ \ref{fig:siteresolved}(c) shows that they are primarily generated inside CoPt and at the CoPt/CuPt interface, while the site-resolved exchange torques show that the magnetization torque acts primarily near at the free surface of CoPt. Large magnetization torque near the free surface of CoPt is also seen for $\tdl$ shown in Fig. \ref{fig:siteresolved}(a); a similar feature was found in calculations for a Co/Pt bilayer \cite{Kirill1}. We note that spin currents generated inside the ferromagnetic layer can induce large dampinglike torques of opposite signs at its opposite surfaces, which have been observed experimentally \cite{Fan_ASOT2019}. The torque at the CoPt/CuPt interface results from the combination of contributions induced by spin-orbit coupling in CoPt and CuPt.

The $3m$ DL torquance is rather small at $V_m=0.65$ eV and falls below the sampling noise at $V_m\geq 0.82$ eV.

\section{Magnetization switching by $3m$ torques}
\label{sec:switching}

Using the analysis of the equations of motion and micromagnetic simulations, this section studies the dynamics of field-free magnetization reversal through domain wall nucleation and propagation driven by $3m$ torques $\tdltm$ and $\tfltm$ in relatively large films~\cite{Liu2021}. We first analytically calculate the pressure on the domain wall, identifying the regimes in which deterministic switching of perpendicular magnetization is possible. The findings of this analysis are validated by micromagnetic simulations performed in the presence of disorder.

\subsection{General Approach}

\begin{figure}
    \centering
    \includegraphics[width=0.9\columnwidth]{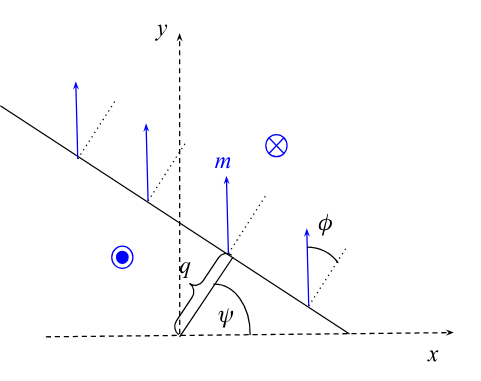}
    \caption{Illustration of collective coordinates $q$,$\psi$ and $\phi$ for the domain wall}
    \label{fig:coordinate}
\end{figure}

To study the dynamics of domain wall motion analytically, we consider a collective coordinate method similar to Ref.~\cite{Boulle2013}. 
A straight domain wall can be described by three variables: its position $q$ , the angle $\phi$ made by the magnetization with the plane of the wall, and the tilting angle $\psi$ made by the normal to the wall and the $x$ axis. This model should be valid whenever the radius of curvature of the domain wall is much greater than the domain wall width. These collective coordinates are illustrated in Fig.~\ref{fig:coordinate}. 

We consider the Lagrangian density of a disorder-free film:
\begin{align}
    \mathcal{L} &= \frac{M_s}{\gamma}(\Dot{\psi}+\Dot{\phi})\cos \theta - A_{ex}(\boldsymbol\nabla \mathbf{m})^2 - K m_z^2 \nonumber\\ 
    &- D \left[m_z (\nabla \cdot \mathbf{m})- (\mathbf{m} \cdot \nabla)m_z)\right] \nonumber\\
    &+ M_s \mathbf{m}\cdot (\mathbf{B}_\mathrm{FL}+\mathbf{B}_{3m\mathrm{FL}}+\mathbf{B}) - \mathcal{E}_{d},
\label{eq:lagrangian}
\end{align}
where $\gamma < 0$, $M_s$, $A_{ex}$, $D$, are the gyromagnetic ratio, saturation magnetization, exchange stiffness and the Dzyaloshinskii–Moriya interaction (DMI), respectively, $\mathbf{B}$ is the external magnetic field, and $\mathcal{E}_d$ is the magnetostatic energy. When the thickness $t$ of the film is sufficiently small, the magnetostatic energy can be taken into account using the effective anisotropy $K= K_u - \mu_0 M_s^2/2$, plus a small correction to the domain wall energy which will be discussed later. $\mathbf{B}_\mathrm{FL}$ and $\mathbf{B}_{3m\mathrm{FL}}$ are the $\mathbf{m}$-dependent effective fields corresponding to the torque terms $\tfl$ and $\tfltm$ in (\ref{expansion}).

To describe the dampinglike terms, it is convenient to introduce the Rayleigh dissipation function
\begin{equation}
    \mathcal{D} = \frac{\alpha M_s}{2 |\gamma|}\left(\Dot{\mathbf{m}}-\frac{\gamma}{\alpha}\mathbf{B}_\mathrm{DL}-\frac{\gamma}{\alpha}\mathbf{B}_{3m\mathrm{DL}}\right)^2.
\label{eq:dissipation}
\end{equation}
which reproduces the correct form of the Gilbert damping and dampinglike torques.
Similar to Section \ref{sec:fp}, we assume the electric current flows in the $\hat{x}$ direction, perpendicular to a mirror plane. The effective fields are:
$\mathbf{B}_\mathrm{FL} = \bfl\,\hat{y}$,
$\mathbf{B}_\mathrm{DL} = \bdl\, \hat{y} \times \mathbf{m}$,
$\mathbf{B}_{3m\mathrm{FL}} = \bfltm(m_y\hat{x} + m_x\hat{y})$,
and $\mathbf{B}_{3m\mathrm{DL}} = \bdltm\mathbf{m}\times (m_y\hat{x} + m_x\hat{y})$. The prefactors for the effective fields are related to the torquances in (\ref{expansion}) as $\bdl=\tdl E/\gamma$, $\bfl=\tfl E/\gamma$, $\bdltm=\sqrt{5/3}\,\tdltm E/\gamma$, and $\bfltm=\sqrt{5/3}\,\tfltm E/\gamma$. Note that the form of the $3m$ effective fields is consistent with Ref.~\cite{Liu2021}, but the component of $\mathbf{B}_{3m\mathrm{FL}}$ parallel to $\mathbf{m}$ has no physical significance, and there is a numerical factor between the definitions of $\bdltm$, $\bfltm$ and the torquances $\tau^{(\nu)}_{2\bar2}$.

Following the collective coordinate method, we write the magnetization in spherical coordinates $\mathbf{m} = \left(\sin \theta\cos\xi, \sin \theta\sin \xi, \cos \theta \right)$ with $\xi=\psi+\phi$ and approximate the domain wall profile as \cite{Boulle2013} $\tan(\theta/2) = \exp[-(x\cos \psi + y\sin \psi-q)/\Delta]$ where
\begin{equation}
\Delta = \sqrt{\frac{A_{ex}}{K_\mathrm{eff}}}
\label{eq:Delta}
\end{equation}
with
\begin{equation}
K_\mathrm{eff}=K-\frac{\pi}{2} M_s \bfl \sin\xi -  M_s \bfltm \sin2\xi,
\label{eq:keff}
\end{equation}
which is the leading-order solution to the minimization of the free energy with respect to the three collective coordinates, along with $\Delta$ as an adiabatic variable. 

We then integrate the Lagrangian and Rayleigh dissipation densities over the system, and obtain the equations of motion for the domain wall from the Euler-Lagrange-Rayleigh equations.

Ignoring contributions due to the curvature of the domain wall, the integrated energy of the domain wall per unit area is then
\begin{align}
    \sigma &= \frac{2A}{\Delta} + 2 K \Delta - \pi D\cos\phi +\mu_0 M_s^2 t\frac{\log 2}{\pi}\cos^2\phi\nonumber\\
    &-\Delta M_s\left(\pi\bfl\sin\xi + 2 \bfltm \sin2\xi\right),
    \label{DWenergy}
\end{align}
where we also accounted for the magnetostatic energy associated with the domain wall \cite{skaugen2019analytical}.
It is convenient to define the pressure responsible for driving the domain wall. The Euler-Lagrange equation for the domain wall position $q$ is
\begin{equation}
    \frac{d}{dt}\frac{\partial \mathcal{L}}{\partial \Dot{q}} = \frac{\partial \mathcal{L}}{\partial q} + \frac{\partial \mathcal{D}}{\partial \Dot{q}},
    \label{eq:Euler-Lagrange}
\end{equation}
where $\partial\mathcal{L}/\partial q$ should include the effective disorder potential, which is not explicitly included in Eq. (\ref{eq:lagrangian}). Assuming the domain wall is pinned, the pressure induced by the magnetic field and current-induced torques is
\begin{align}
    P = M_s\left( B_z - \frac{\pi}{2} \bdl\cos \xi  + \bdltm \cos 2 \xi\right).
\end{align}

Note that the pressure induced by the conventional dampinglike torque is proportional to $m_x=\cos\xi$. In a large, multi-domain sample with sufficiently smooth disorder potential, the switching process is primarily controlled by the pressure $P$. The non-trivial feature is that its current-induced part, in contrast to the field-induced pressure, depends on the orientation of the domain wall.

\subsection{Pressure on a N\'eel Wall}

\label{sec:Neel}

In thin films with substantial DMI, it is typical that $\pi D \gg \mu_0 M_s^2 t \frac{\log2}{\pi}$, and the magnetostatic contribution to the DW energy may be ignored. In this section we consider the pressure on a domain wall in the limit $t\rightarrow 0$, and the implications for switching by the $3m$ fieldlike torque in these systems.

Using the method outlined in the previous section for an infinite system, $\Dot{\psi} \rightarrow 0$ as the domain wall length goes to infinity, and assuming $\Dot{\phi} = 0$ the equations of motion become
\begin{align}
     -\frac{\alpha}{\gamma} \frac{\Dot{q}}{\Delta} + T^\mathrm{DL}_\mathrm{eff}(\xi) 
    &= B_z ,\label{eq:eom1} \\
    \frac{\pi D}{2\Delta M_s}\sin\phi - \mu_0 M_s \frac{t\log 2}{2\pi \Delta}\sin 2\phi &= T^\mathrm{FL}_\mathrm{eff}(\xi) \label{eq:eom2}.
\end{align}
Here
\begin{align}
    T^\mathrm{FL}_\mathrm{eff}(\xi)&=\frac{\pi}{2} (\bfl\cos\xi+B_x\sin \xi) + 2 \bfltm \cos2\xi,\\
    T^\mathrm{DL}_\mathrm{eff}(\xi)&=\frac{\pi}{2} \bdl\cos \xi + \bdltm \cos2\xi,
\end{align}
are the collective torques acting on the domain wall that appear due to fieldlike and dampinglike spin torques.

Equation (\ref{eq:eom1}) describes the wall's response to the applied pressure for fixed $\psi$ and $\phi$, while equation (\ref{eq:eom2}) gives a transcendental equation for $\phi$ at fixed $\psi$. In particular, as $t\rightarrow 0$
\begin{equation}
    \frac{ \sin\phi}{k_\text{N\'eel}\left(\cos{\left(\psi + \phi\right)} + \delta \cos{2\left(\psi + \phi\right)} \right)}=1,
\end{equation}
where $k_\text{N\'eel} = \Delta M_s \bfl/D$ and $\delta = \frac{4\bfltm}{\pi\bfl}$. This equation has no simple analytical solution, so we expand in terms of $\delta \ll 1$ and look for perturbative solutions of the form $\phi = \phi_0(\psi) + \delta \phi_1(\psi)$. The zeroth order solutions are
\begin{equation}
    \phi_0 = \cot^{-1} \left(\frac{1}{k_{\text{N\'eel}} \cos\psi} +\tan \psi \right) + n\pi,\quad n\in \mathbb{Z}.
    \label{eq:neel0}
\end{equation}

For any given $\psi$, there are two unique solutions, 
one of which corresponds to the energy minimum of the domain wall, and the other to the maximum. In the limit of small current density $k_{\text{N\'eel}} \ll 1$ the solutions tend to $0$ and $\pi$ describing two N\'eel-like domain wall solutions. As the current density increases, the magnetization inside the domain wall tends to rotate away from the N\'eel configuration ($\xi=0$ or $\pi$) toward $\mathbf{B}_\mathrm{FL}$. The solutions undergo a phase transition near the critical point $k_{\text{N\'eel}}=1$ where the in-plane moment of the domain wall aligns close to the direction of the effective field of the spin orbit torques, regardless of the orientation of the wall. 

\begin{figure}
    \centering
    \includegraphics[width=0.9\columnwidth]{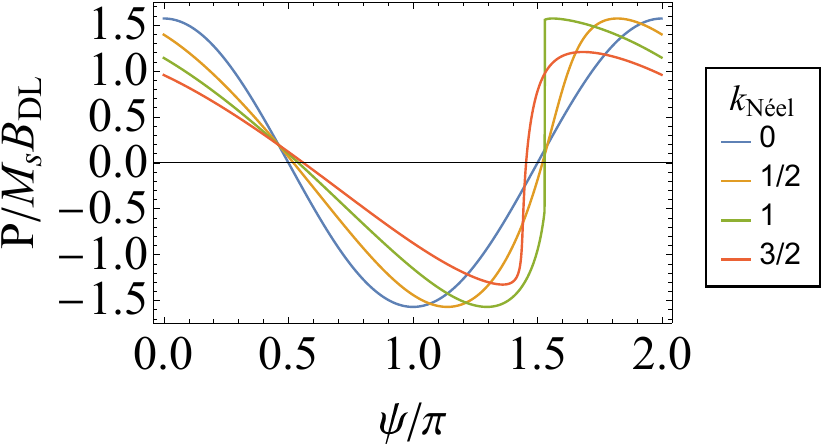}
    \caption{Normalized pressure $P/M_s\bdl$ for a N\'eel-like domain wall for several values of $k_{\text{N\'eel}}$, and $\delta=0.1$.}
    \label{fig:PpsiNeel}
\end{figure}

The correction $\phi_1$ can be written in the form
\begin{equation}
    \phi_1 = \pm \frac{ \sec \psi  \cos (\psi +\phi_0) (k_{\text{N\'eel}} (k_{\text{N\'eel}}+2 \sin \psi )-\cos 2 \psi )}{k_{\text{N\'eel}}+2 \sin \psi +1/k_{\text{N\'eel}}},
    \label{eq:neel1}
\end{equation}
where the $\pm$ sign is determined by the branch chosen in (\ref{eq:neel0}). We use these results to plot the normalized pressure $P/M_s\bdl$ as a function of $\psi$ for different $k_{\text{N\'eel}}$ with fixed $\delta$ in Fig. \ref{fig:PpsiNeel}.

\begin{figure}
    \centering
    \includegraphics[width=0.9\columnwidth]{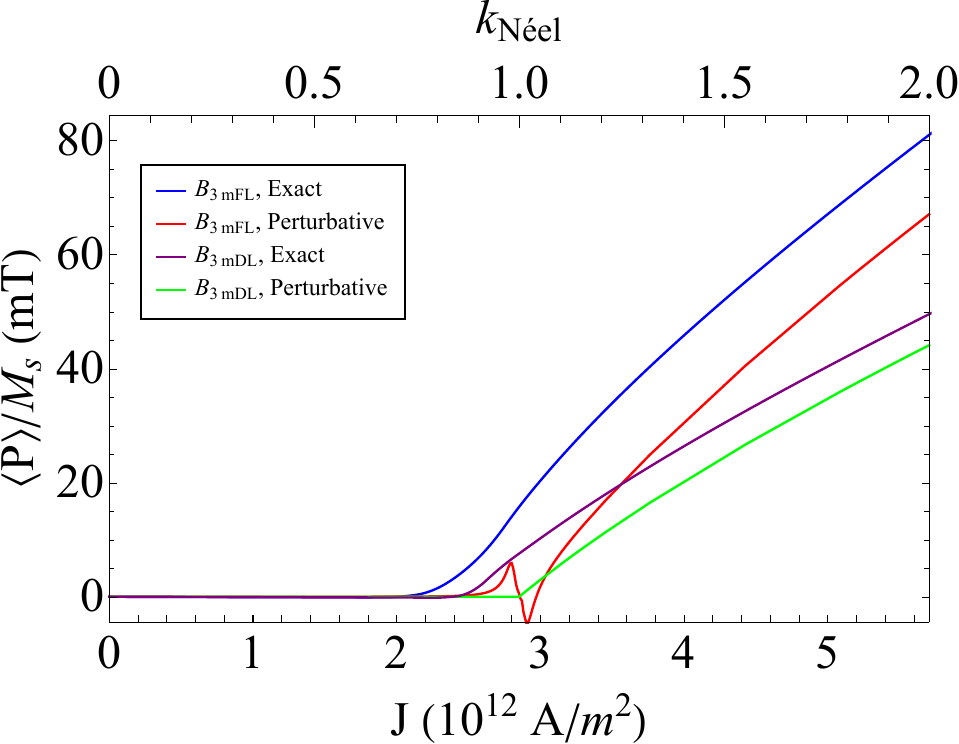}
    \caption{Pressure averaged over $\psi$ for material parameters: $M_s = 850\times 10^3$ A/m, $K = 1 $ MJ/m\textsuperscript{3}, $A_{ex} = 17$ pJ/m, $D = 1$ mJ/m\textsuperscript{2}, $\bdl=\bfl= 0.1$ T/($10^{11}$A/m\textsuperscript{2}), $\delta = 0.1$. (a) the exact numerical solution for $\bfltm$, (b) the perturbative approximation for $\bfltm$, (c) the exact numerical solution for $\bdltm = 0.1 \bdl$, (d) the perturbative result for $\bdltm = 0.1 \bdl$.}
    \label{fig:PavgNeel}
\end{figure}

In order to estimate the efficiency of switching by the $3m$ fieldlike torque compared to an out of plane field we consider pressure averaged over $\psi$, $\langle P \rangle = 1/2\pi \int \frac{\pi}{2}M_s\bdl \cos\xi d\psi$, shown in Fig. \ref{fig:PavgNeel}. A positive pressure corresponds to expansion of the $m_z = -1$ downward pointing domains, while negative pressure results in expansion of $m_z = +1$ domains. For small $k_{\text{N\'eel}}$ the pressure is nearly antisymmetric after rotation by angle $\pi$, resulting in translational motion rather than net expansion of a domain, consistent with the usual method of driving N\'eel bubble domains or skyrmions by damping-like SOT. 

As seen in Fig. \ref{fig:PavgNeel}, the average pressure is essentially zero until the current density approaches the critical point near $k_{\text{N\'eel}} = 1$. For the first order approximation in $\delta$ this phase transition occurs exactly at $k_{\text{N\'eel}} = 1$. Finite values of $\delta$ decrease the value of the critical point, but will always correspond to a value of $k_{\text{N\'eel}}>0.5$ if $\bfltm < 
\bfl$. This suggests a relationship between the applied current density and the possibility of deterministic switching in systems with N\'eel domain walls; $|J| \gtrsim \rho \gamma D/\Delta M_s \tau_{FL}$ where $\rho$ is the electrical resistivity. For typical material parameters this will require current density larger than $10^{12}$ A/m\textsuperscript{2}.
At the critical point the effective field from the SOT exceeds the effective field of the DMI, and the domain wall is pulled away from a N\'eel-like configuration, and $\langle P \rangle$ becomes non-zero. For $k_{\text{N\'eel}} \gg 1$ the average pressure becomes $\langle P \rangle = \frac{\pi}{2}\delta M_s\bdl$, which is linear in the current density. 

For 3m damping-like torque without 3m field-like torque, the first order correction to $\phi$ is zero, leading to $\langle P \rangle = 0$, but the pressure contains an additional term $\langle P_{3m} \rangle = 1/2\pi \int M_s\bdltm \cos 2\xi d\psi$. This average pressure shows similar qualitative behavior, but with a decreased slope in the linear regime for the same field value.

Thus, a very large current density is required to achieve a non-zero average pressure on a N\'eel-type domain wall. In the next section, we show that the situation is more favorable for switching in systems with weak DMI where the domain walls are of the Bloch type.

\subsection{Pressure on a Bloch Wall}

\label{sec:Bloch}

Next we will consider the case where DMI is weak, $\pi D \ll \mu_0 M_s^2t\frac{\log 2}{\pi}$.
In this case the transcendental equation for $\phi$ is
\begin{equation}
    \frac{\sin2\phi}{k_{\text{Bloch}}\left(\cos(\psi+\phi) + \delta \cos2(\psi+\phi)\right)} = -1,
    \label{eq:PhiBloch}
\end{equation}
where $k_{\text{Bloch}} = \pi^2 \Delta \bfl /  2\mu_0 t M_s \log 2$. This equation can be solved numerically or perturbatively, treating both $k_{\text{Bloch}}$ and $\delta$ as small parameters. Without spin orbit torques, two solutions of Eq. (\ref{eq:PhiBloch}) with $\phi = \pm \pi/2$ correspond to energy minima describing two degenerate Bloch states, while solutions $\phi = 0,\pi$ correspond to energy maxima. When the torques are finite but small ($k_{\text{Bloch}} \ll 1$), the degeneracy is lifted, and the energy difference between the two Bloch-like states is linear in $k_{\text{Bloch}}$ and depends on $\psi$. The global minimum at a given $\psi$ corresponds to the Bloch-like wall with a positive projection of the magnetization on $\mathbf{B}_\mathrm{FL}$. The global minimal energy configuration then has Bloch lines near $\psi = \pm \pi/2$. For a large domain, the increase in energy associated with these Bloch lines should be small compared to the decrease in energy by aligning the magnetic moments inside the DW with $\mathbf{B}_\mathrm{FL}$. 

\begin{figure}
    \centering
    \includegraphics[width=0.9\columnwidth]{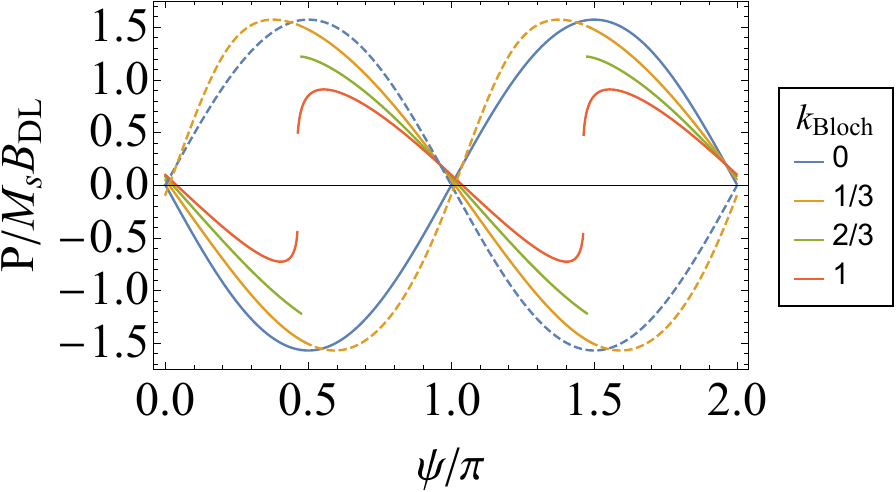}
    \caption{Normalized pressure $P/M_s\bdl$ for a Bloch-like domain wall for several values of $k_{\text{Bloch}}$, with $\delta=0.1$. Solid lines represent the lowest energy state; dashed lines the second lowest energy state.}
    \label{fig:PpsiBloch}
\end{figure}

The normalized pressure $P/M_s\bdl$ for both local minima is plotted as a function of $\psi$ in Fig. \ref{fig:PpsiBloch}. For this choice of sign for $k_{\text{Bloch}}$ and $\delta$, an $m_z = -1$ domain should expand mostly from the regions $\psi \in (\pi/2, \pi)$ and $\psi \in (3\pi/2, 2\pi)$, and collapse at a slightly slower speed from the regions $\psi \in (0, \pi/2)$ and $\psi\in (\pi, 3\pi/2)$, leading to a positive average pressure.

\begin{figure}
    \centering
    \includegraphics[width=0.9\columnwidth]{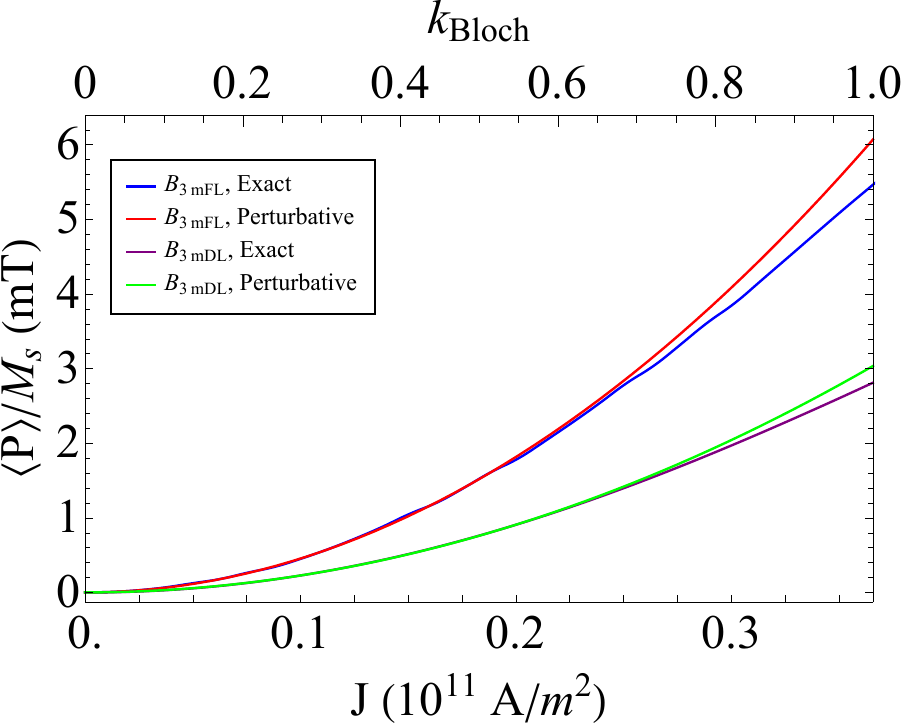}
    \caption{Pressure averaged over $\psi$ for material parameters: $M_s = 850\times 10^3$ A/m, $K = 1 $ MJ/m\textsuperscript{3}, $A_{ex} = 17$ pJ/m, $t = 1$ nm, $\bdl=\bfl= 0.1$ T/($10^{11}$A/m\textsuperscript{2}), $\delta = 0.1$. (a) the exact numerical solution for $\bfltm$, (b) the approximation (\ref{eq:BlochPressure}), (c) the exact numerical solution for $\bdltm$, (d) the perturbative result for $\bdltm$.}
    \label{fig:PavgBloch}
\end{figure}

We show the average pressure $\langle P \rangle$ as a function of $k_{\text{Bloch}}$ and current in Fig. \ref{fig:PavgBloch}, assuming that $\phi$ lies in the global minimum for all $\psi$, using both perturbative and numerical methods. To second order in the current density the perturbative solution is 
\begin{equation}
    \langle P \rangle = \frac{3\pi\delta}{2e}k_{\text{Bloch}}M_s\bdl = \frac{6\pi^2\Delta \bfltm \bdl}{e \ln(2)  \mu_0 t},
    \label{eq:BlochPressure}
\end{equation}
where $e$ is the base of the natural logarithm. This gives excellent agreement with the numerical result when $\delta=0.1$.

We see nearly quadratic dependence of the pressure on the current density, which allows for the possibility of deterministic switching of perpendicular magnetization at relatively small current densities, in contrast to the N\'eel case.

At small current densities it may be important to consider the role of the other local minimum in the pressure calculation. For $k_{\text{Bloch}} < 1/2$ the average pressure of this state is exactly opposite to that of the global minimum. This may lead to a decrease in the overall pressure if some segments of the wall settle in the unfavorable local minimum, however this effect should vanish as the SOT increase and the energy difference between the global and local minimum becomes sufficiently large. 

Near $k_{\text{Bloch}}>1/2$ the domain wall undergoes a phase transition where the energy of the unfavorable Bloch state crosses the energy of the N\'eel states for certain ranges of angles $\psi$. In this case the global minimum remains a perturbed Bloch-like state, but the second minimum is a hybridization of Bloch and N\'eel like states depending on domain wall orientation.

The pressure due to 3m damping-like torque without 3m field-like torque is again similar to the result for 3m field-like torque; The average pressure $\langle P \rangle = 0$, and $\langle P_{3m} \rangle= \frac{3\pi^2\Delta \bdltm \bdl}{e \ln(2)  \mu_0 t}$ giving a total pressure of exactly one half the result for 3m field-like torque, for the same field values.

When the current density is large enough that $k_{\text{Bloch}} \geq 1$, the effective fields of the spin orbit torques exceed the in-plane component of the demagnetizing field, and the domain wall cannot be accurately described as a perturbed Bloch wall. In the limit that $k_{\text{Bloch}} \gg 1$ the average pressure again becomes $\langle P \rangle = \frac{\pi}{2}\delta M_s\bdl$, the same as in the N\'eel case.

\subsection{Micromagnetic Simulations of Magnetization reversal by $3m$ Torque} 

The treatment in Sections \ref{sec:Neel} and \ref{sec:Bloch} is based on the equations of motion for the collective coordinates of a straight domain wall. This treatment is expected to work well in the flow regime, when disorder is effectively weak, and curvature of the domain wall is negligible. Such treatment may be inapplicable in the creep regime or near the depinning transition for strong disorder where the domain wall can effectively adapt to the disorder potential. To investigate different regimes and identify crossovers, we turn to micromagnetic simulations of current-induced magnetization switching based on the integration of the Landau-Lifshitz-Gilbert (LLG) equation in the presence of disorder. These simulations were performed using the MuMax3 code \cite{mumax3}.

In our micromagnetic simulations, we consider a ferromagnetic square of width $\SI{500}{\nano\meter}$ ($512\times512$ unit cells) with periodic boundary conditions, in the presence of current-induced SOT with the $3m$ FL component. The thickness is $\SI{1}{\nano\meter}$. The material parameters are the following: saturation magnetization $M_s=\SI{850}{\kilo\ampere\per\meter}$, exchange stiffness $A_{ex}=\SI{17}{\pico\joule\per\meter}$, uniaxial anisotropy constant $K_u=\SI{1}{\mega\joule\per\cubic\meter}$, and damping constant $\alpha=0.5$. The large value of $\alpha$ was chosen to improve convergence and reduce run time of the simulations, but our results do not qualitatively depend on this value. We assume $\bdl = \bfl = 0.1 $ T/($10^{11}$A/m\textsuperscript{2}), $\bfltm = 0.1 \bfl$. All simulations were performed at a temperature $T = 50$ K.

Disorder was introduced by creating grains with average size of 8 nm, with anisotropy chosen from a random normal distribution centered on $K_u$ with a width $\sigma_{K}$ of 10\%.

First we concentrate on a system which prefers N\'eel domain walls. Here we introduce a DMI of $D = 1$ mJ/m\textsuperscript{2}, so that the DMI contribution to the domain wall energy is much greater than the magnetostatic part. 

\begin{figure}
    \centering
    \includegraphics[width=1\columnwidth]{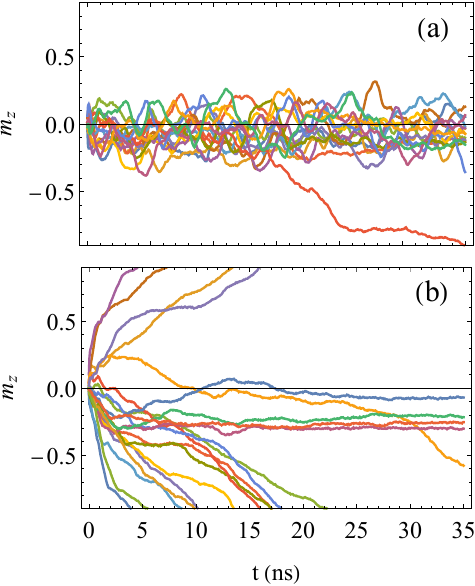}
    \caption{Micromagnetics results including 3m field-like torque, for $m_z$ vs time starting from many different random configurations for (a) N\'eel-like domain walls for current density of $J=2.5\times 10^{10}$ A/m\textsuperscript{2} (b) Bloch-like domain walls for $J=2\times10^{10}$ A/m\textsuperscript{2}. $M_s=\SI{850}{\kilo\ampere\per\meter}$, $A_{ex}=\SI{17}{\pico\joule\per\meter}$,  $K=\SI{0.5}{\mega\joule\per\cubic\meter}$, $\alpha=0.5$, $\bdl = \bfl = 0.1 $ T/($10^{11}$A/m\textsuperscript{2}), $\bfltm = 0.1 \bfl$, $T = 50$K.}
    \label{fig:mzvt3mFL}
\end{figure}

\begin{figure}
    \centering
    \includegraphics[width=1\columnwidth]{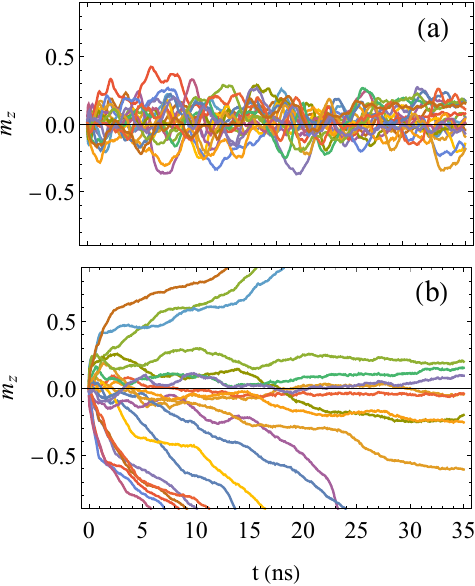}
    \caption{Micromagnetics results for $m_z$ vs time including 3m damping-like torque, starting from many different random configurations for (a) N\'eel-like domain walls for current density of $J=2.5\times 10^{10}$ A/m\textsuperscript{2} (b) Bloch-like domain walls for $J=2\times10^{10}$ A/m\textsuperscript{2}. $M_s=\SI{850}{\kilo\ampere\per\meter}$, $A_{ex}=\SI{17}{\pico\joule\per\meter}$,  $K=\SI{0.5}{\mega\joule\per\cubic\meter}$, $\alpha=0.5$, $\bdl = \bfl = 0.1 $ T/($10^{11}$A/m\textsuperscript{2}), $\bdltm = 0.1 \bfl$, $T = 50$K.}
    \label{fig:mzvt3mDL}
\end{figure}

The initial configuration of the system is obtained by using the MuMax3 minimize function, beginning with a random configuration. Then the torques are enabled with a constant current density, and the simulations are run for $30$ ns. Figures \ref{fig:mzvt3mFL}(a) and \ref{fig:mzvt3mDL}(a) show that the final magnetization $m_z$ is nearly random for current densities less than $J \approx 3\times 10^{11}$ A/m\textsuperscript{2}. In this regime, we see a period of initial relaxation, after which domains are driven across the system in a nearly uniform manner. For greater values of $J$, we nearly always see successful magnetization reversal. This is in rough agreement with the analysis in Section \ref{sec:Neel}, as a value of $k_{\text{N\'eel}} =1 $ corresponds to a current density of $J = 6 \times 10^{11}$ A/m\textsuperscript{2}.

Next, we consider magnetization reversal in systems with Bloch-like domain walls. The simulation parameters in this case are identical to the previous discussion, except now $D = 0$. Figure \ref{fig:mzvt3mFL}(b) and \ref{fig:mzvt3mDL}(b) show that a final configuration with $m_z < 0$ is more likely than a state with $m_z >0$, even for relatively small $J$. 

We also note that the domains realized in micromagnetics do not always correspond to the true energy minimum described in Section \ref{sec:Bloch}. This is primarily due to the random initial configuration, and the fact that it takes less time for a system of this size to reverse completely than for the domain wall to relax to the energy minimum.

\begin{table}
    \begin{tabular}{c|c|c}
    \hline
         & N\'eel & Bloch  \\
    \hline
     $J$ (10\textsuperscript{10} A/m\textsuperscript{2}) & 2.5 & 2.0 \\
       $D$ (mJ / m\textsuperscript{2}) & 1.0 & 0 \\
    \hline
       $A_{ex}$ (pJ m\textsuperscript{-1})  & \multicolumn{2}{c}{17} \\
       $M_s$ (kA m\textsuperscript{-1}) & \multicolumn{2}{c}{850} \\
       $K_u$ (MJ m\textsuperscript{-3}) & \multicolumn{2}{c}{1.0} \\
       $\sigma_{K}$ (MJ m\textsuperscript{-3}) & \multicolumn{2}{c}{0.1}\\
       $\alpha$ & \multicolumn{2}{c}{0.5} \\
       $\bdl/J$ (T/(10\textsuperscript{11}A/m\textsuperscript{2})) & \multicolumn{2}{c}{0.1} \\
       $T$ (K) & \multicolumn{2}{c}{50} \\
    \hline
      
    \end{tabular}
    \caption{Table of parameters used in micromagnetics in Figs \ref{fig:mzvt3mFL} and \ref{fig:mzvt3mDL}. In all simulations $\bdl = \bfl$. In simulations with the 3m field-like torque $\bfltm = 0.1 \bfl$ and $\bdltm = 0$. In simulations with 3m damping-like torque $\bdltm = 0.1 \bfl$ and $\bfltm=0$.}
\end{table}

\section{Conclusions}
Motivated by a recent experiment~\cite{Liu2021}, we have studied spin-orbit torques and magnetization reversal in an epitaxial L1$_1$-ordered CoPt/CuPt bilayer system with perpendicular magnetic anisotropy and $3m$ ($C_{3v}$) symmetry. The torques are represented in terms of an expansion in vector spherical harmonics~\cite{Kirill2}, and symmetry analysis is used to identify all allowed terms for a system with $C_{nv}$ symmetry for an arbitrary $n$. Using first-principles nonequilibrium Green’s function formalism combined with Anderson disorder, we have found sizable spin-orbit torques enabled in the CoPt/CuPt bilayer by the $C_{3v}$ symmetry.

Using the values of fieldlike, dampinglike, and $3m$ spin-orbit torques obtained in the first-principles calculations as a guide, we have numerically and analytically studied the feasibility of perpendicular magnetization reversal in a film of a large lateral size~\cite{Liu2021}, assuming that switching proceeds through domain wall motion in a multi-domain state. Such switching can only be deterministic in the presence of symmetry breaking introduced by the $3m$ torques or in-plane magnetic field. We calculate the pressure acting on a domain wall as a function of its spatial orientation. We observe that the current-induced torques affect the relative stability of different internal configurations of the domain wall. As a result, in the analytical calculations we assume, somewhat unrealistically, that the configuration with the lower energy is always selected for each orientation. This preference provides a mechanism for deterministic switching of perpendicular magnetization assisted by $3m$ torque. 

We find that for systems with strong interfacial DMI characterized by the N\'eel character of domain walls, a very large minimal current density is required to achieve deterministic switching of perpendicular magnetization. This is because switching in this regime requires a large reorientation of the magnetization inside the domain walls away from the N\'eel configuration.

\begin{figure}[htb]
    \centering
    \includegraphics[width=0.9\columnwidth]{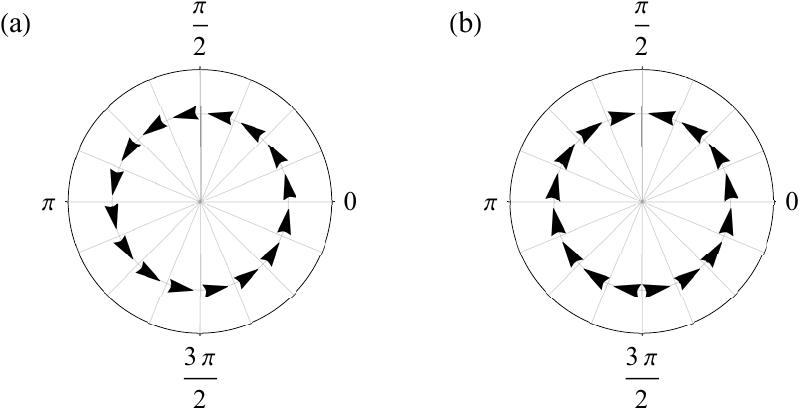}
    \caption{Plot of the in-plane component of the magnetization for the minimum energy configuration of a Bloch domain wall as a function of angle $\psi$ for (a) $J = 0$, (b) In the limit as $J \rightarrow 0$ from above the magnetization reverses direction across the lines $\psi = \pi/2$ and $\psi = 3\pi/2$, increasing the projection onto $\mathbf{\bfl}$, and minimizing the domain wall energy. }
    \label{fig:bloch_minimum}
\end{figure}

For films with relatively weak interfacial DMI the domain walls have Bloch-type structure. For such walls, even a small tilting of the magnetization inside the domain wall is, in principle, sufficient for deterministic switching. 
However, the current-induced pressure acting on the domain wall depends strongly on its orientation, and its orientation average depends quadratically on the current density, exhibiting non-analytical behavior at zero current due to the branching point between the energies of two domain wall chiralities, shown in Fig.~\ref{fig:bloch_minimum}.

Thus, we find that the Bloch-type wall case is more favorable, compared to N\'eel-type, for deterministic switching assisted by the fieldlike $3m$ torque. However, even in this case the quadratic scaling of the average driving force with the current density is unfavorable for applications. We observe similar behavior due to the dampinglike $3m$ torque. We also note that in a real sample the internal domain wall configurations may not correspond to the lowest energy and be subject to hysteresis. Such deviations may make deterministic switching even more difficult to achieve. This is in contrast to switching in the presence of an in-plane magnetic field, where linear scaling with current density is possible for both N\'eel- and Bloch-type domain walls.

Our results suggest that studies of deterministic switching assisted by crystallographic spin-orbit torques should focus on systems with the Bloch-type domain walls.

\acknowledgements

We are grateful to Fei Xue and Paul Haney for useful discussions.
This work was supported by the National Science Foundation through Grant No. DMR-1916275 and DMR-2324203. AAK and ES were supported by the U.S. Department of Energy, Office of Science, Basic Energy Sciences, under Award No. DE-SC0021019. 
Calculations were performed utilizing the Holland Computing Center of the University of Nebraska, which receives support from the Nebraska Research Initiative.

\appendix

\section{Symmetry analysis}
\label{app:symmetry}

The torque \textbf{T} induced by the applied electric field \textbf{E} is given by torquance tensor $\hat{K}(\hat{m})$,
\begin{equation}
    \textbf{T}=\hat{K}(\hat{m})\textbf{E},
\end{equation}
and it can be expanded in the VSH basis as
\begin{equation}
    \hat{K}(\hat{m})=\sum_{lm\nu}\textbf{Y}_{lm}^{(\nu)}\otimes \textbf{K}_{lm}^{(\nu)},
\end{equation}
where $\textbf{Y}_{lm}^{(\nu)}$ are VSH and $\textbf{K}_{lm}^{(\nu)}$ are complex Cartesian vectors which we write as follows: $\textbf{K}_{lm}^{(\nu)}=K_{lm,+}^{(\nu)}(\hat{x}+i\hat{y})+K_{lm,-}^{(\nu)}(\hat{x}-i\hat{y})$. We have aligned the $z$ axis perpendicular to the film plane, and we will eventually assume the $x$ axis is chosen parallel to $\mathbf{E}$. Because the torquance tensor is real, there is a restriction on the coefficients:
$K_{lm,+}^{(\nu)}=(-1)^m K_{l,-m,-}^{(\nu)*}$.

If the symmetry group of the film includes a $C_{n}$ axis, the torquance tensor $\hat{K}(\hat{m})$ should be invariant with respect to the rotation by angle $\alpha=2\pi/n$ around the $z$ axis. A VSH transforms as $\mathbf{Y}_{lm}^{(\nu)}(\theta,\phi)\rightarrow e^{im\alpha}\mathbf{Y}_{lm}^{(\nu)}(\theta,\phi)$ and Cartesian vectors transform as $\mathbf{K}_{lm}^{(\nu)}\rightarrow e^{i\alpha}K_{lm,+}^{(\nu)}(\hat{x}+i\hat{y})+e^{-i\alpha}K_{lm,-}^{(\nu)}(\hat{x}-i\hat{y})$. We then require that the torquance tensor is invariant:
\begin{align}
    &e^{im\alpha}\mathbf{Y}_{lm}^{(\nu)}[e^{i\alpha}K_{lm,+}^{(\nu)}(\hat{x}+i\hat{y})+e^{-i\alpha}K_{lm,-}^{(\nu)}(\hat{x}-i\hat{y})]\nonumber\\
    &=\mathbf{Y}_{lm}^{(\nu)}[K_{lm,+}^{(\nu)}(\hat{x}+i\hat{y})+K_{lm,-}^{(\nu)}(\hat{x}-i\hat{y})]
    \label{invariance}
\end{align}
where we only kept the relevant in-plane components. Equation (\ref{invariance}) should be satisfied as an identity. Because $\hat{x}\pm i\hat{y}$ are linearly independent vectors, in the case $e^{i\alpha}\neq e^{-i\alpha}$ (i.e., $n\neq2$) it must be satisfied individually for the $\hat{x}+i\hat{y}$ and $\hat{x}-i\hat{y}$ terms. Thus, with $n\geq3$, nonzero $K_{lm,\pm}^{(\nu)}$ requires $e^{i(m\pm1)2\pi/n}=1$, or $m=ns\mp 1$, where $s$ is an arbitrary integer.

The case $n=2$ is special in that (\ref{invariance}) is satisfied identically at any odd $m$, which can also be written formally as $m=ns\pm1$ but with no restrictions on $K_{lm,\pm}^{(\nu)}$.

Now consider the effect of the symmetry planes included in the $C_{nv}$ symmetry group. If there is a $\sigma_x$ mirror plane, its effect is the following (taking into account that both $\hat m$ and the torque are axial vectors): $\mathbf{Y}_{lm}^{(\nu)}(\theta,\phi)\rightarrow (-1)^l\mathbf{Y}_{l-m}^{(\nu)}(\theta,\phi)$ and $\mathbf{K}_{lm}^{(\nu)}\rightarrow -K_{lm,-}^{(\nu)}(\hat{x}+i\hat{y})-K_{lm,+}^{(\nu)}(\hat{x}-i\hat{y})$. The torquance tensor in response to an electric field along the $x$ direction should then be proportional to $\mathbf{Y}_{lm}^{(\nu)}-(-1)^{l+m}\mathbf{Y}_{lm}^{(\nu)*}$. When $l+m$ is even (odd), this symmetry-allowed term is the imaginary (real) part of the complex VSH $\mathbf{Y}_{lm}^{(\nu)}$, which is proportional to the real VSH $\mathbf{Z}_{l,m}^{(\nu)}$ with $m<0$ ($m>0$). Thus, for the $C_{nv}$ symmetry group including a $\sigma_x$ mirror plane, the allowed real VSH are those with $m=ns\pm 1 < 0$ for even $l+m$ and with $m=ns\pm 1 > 0$ for odd $l+m$.

If instead the crystal is aligned such that one of the $C_{nv}$ symmetry planes is $\sigma_y$, a similar analysis shows that the allowed real VSH (for $\mathbf{E}\parallel\hat x$) are those with $m>0$ ($m<0$) for even (odd) $l$. We note that the resulting condition for $\sigma_x$ and $\sigma_y$ is the same for odd $m$ and opposite for even $m$.

Table \ref{tab:vsh} lists the allowed real VSH for the $C_{3v}$ symmetry group.

\begin{table}[htb]
\caption{Symmetry-allowed real VSH for a system with the $C_{3v}$ ($3m$) symmetry group and $\mathbf{E}\parallel\hat x$.}

\begin{tabular}{ll}
\hline
\multirow{2}{*}{$\mathbf{E} \perp$ mirror plane}  &  $\mathbf{Z}_{1,-1}^{(\nu)}$, $\mathbf{Z}_{2,1}^{(\nu)}$, $\mathbf{Z}_{3,-1}^{(\nu)}$, $\mathbf{Z}_{4,1}^{(\nu)}$ $\cdots$ \\
 & $\mathbf{Z}_{2,-2}^{(\nu)}$, $\mathbf{Z}_{3,2}^{(\nu)}$, $\mathbf{Z}_{4,-2}^{(\nu)}$, $\mathbf{Z}_{5,2}^{(\nu)}$ $\cdots$\\
\hline     
\multirow{2}{*}{$\mathbf{E} \parallel$ mirror plane}  &  $\mathbf{Z}_{1,-1}^{(\nu)}$, $\mathbf{Z}_{2,1}^{(\nu)}$, $\mathbf{Z}_{3,-1}^{(\nu)}$, $\mathbf{Z}_{4,1}^{(\nu)}$ $\cdots$ \\
 & $\mathbf{Z}_{2,2}^{(\nu)}$, $\mathbf{Z}_{3,-2}^{(\nu)}$, $\mathbf{Z}_{4,2}^{(\nu)}$, $\mathbf{Z}_{5,-2}^{(\nu)}$ $\cdots$\\
\hline
\end{tabular}
\label{tab:vsh}
\end{table}

\section{Site-resolved torques}
\label{app:siteres}

\begin{figure*}[htb]
\includegraphics[width=1\textwidth]{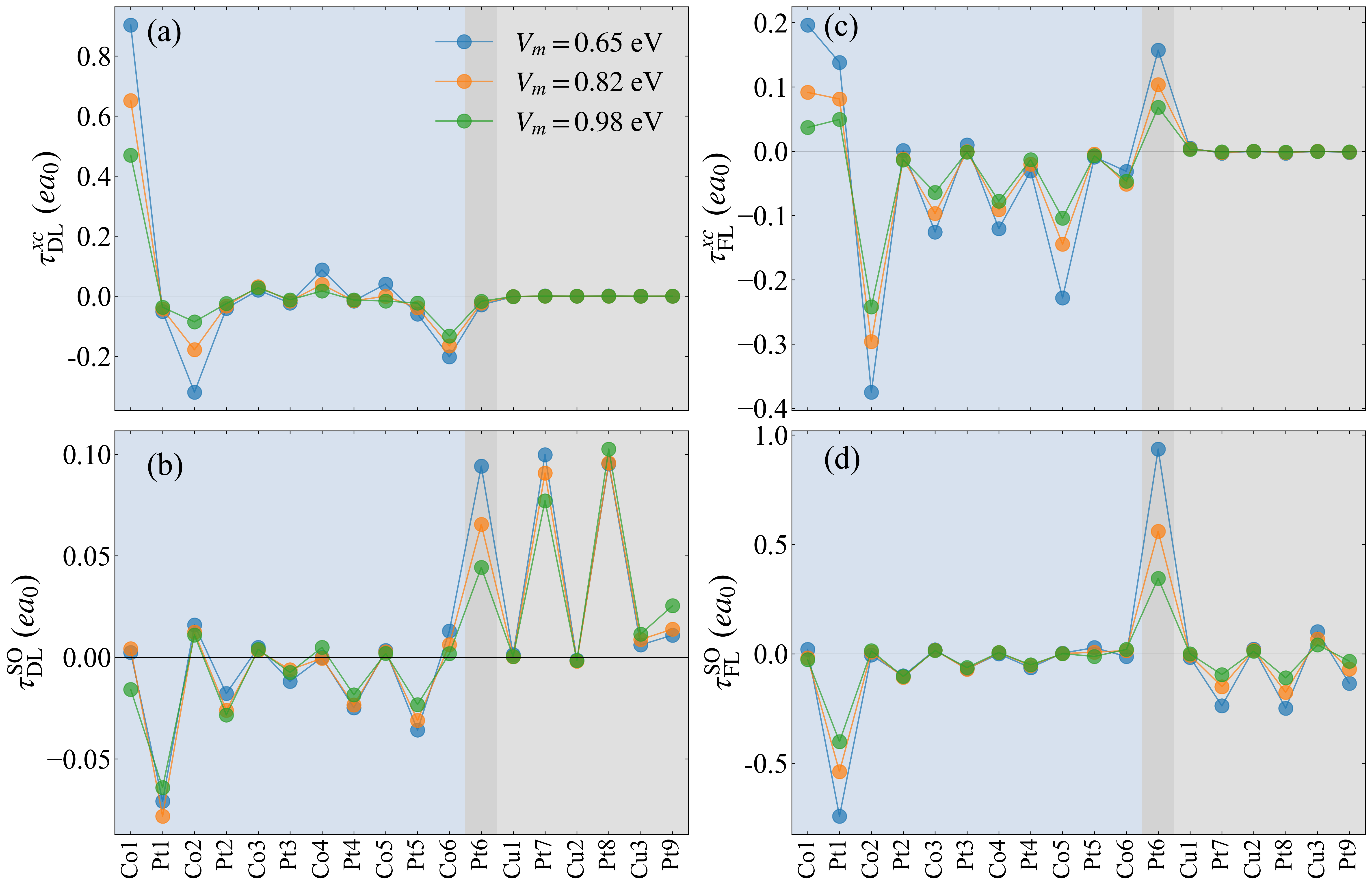}
\caption{Disorder dependence of site-resolved torquances. (a) $\tau^{xc}_\mathrm{DL}$ (b) $\tau^\mathrm{SO}_\mathrm{DL}$ (c) $\tau^{xc}_\mathrm{FL}$ (d) $\tau^\mathrm{SO}_\mathrm{FL}$. The Fermi sea contribution to $\tdl$ is not included.}
\label{fig:FL-site}
\end{figure*}

Figure \ref{fig:FL-site} shows the disorder dependence of DL and FL SOT. For $\tau^{xc}_\mathrm{DL}$, the torque on the Co layer at the free surface (Co1) depends strongly on the disorder strength while the one near the interface (Co6) depends weakly on the disorder strength. For $\tau^{xc}_\mathrm{FL}$, the torque on each Co layer depends strongly on the disorder strength. For $\tau^{SO}_\mathrm{DL}$, the spin current is generated at each Pt layer and the Pt layer near the interface (Pt6) depends strongly on the disorder strength. For $\tau^{xc}_\mathrm{FL}$, the spin currents are mainly generated at the two Pt layers near the free surface (Pt1) and interface (Pt6). They have opposite contributions to the torque and decrease with increasing disorder strength.

\section{Calculation of the Average Pressure}

We calculate the average pressure using

\begin{equation}
    \langle P \rangle = \frac{1}{2\pi}\int_0^{2\pi}M_s\bdl \cos{(\psi +\phi)}d \psi
\end{equation}

We then expand to first order in $\delta$, $\phi = \phi_0 + \delta \phi_1$ and drop the term  $\cos{\psi +\phi_0}$ which integrates to zero. We are left with;

\begin{equation}
    \langle P \rangle \approx -\frac{1}{2\pi}\int_0^{2\pi}\delta M_s\bdl  \phi_1 \sin{(\psi + \phi_0)} d\psi
    \label{eq:pC2}
\end{equation}

This shows the leading contribution to the pressure must come from the term $\delta \phi_1 \sin{(\psi + \phi_0)}$.

In the N\'eel case when $J\ll 1$ we have to leading order $\phi_0 \approx 0$ or $\pi$ depending on the sign of $D$, and $\phi_1 \approx -k_{\text{N\'eel}}\cos 2\psi$ and therefore the average pressure $\langle P \rangle \propto \int \cos 2\psi \sin\psi = 0$ to second order in the current density (since $\bdl$ is also proportional to the current density). 

For Bloch domain walls when $J\ll 1$, we can simplify to $\phi_0 = \pm \pi/2 \mp \frac{1}{2}k_{\text{Bloch}} \sin\psi$, and $\phi_1 = \frac{1}{2}k_{\text{Bloch}} (-1\pm(1+\frac{1}{2}\cos2\psi))$, where the upper (lower) sign corresponds to the energy minimum in the interval $\psi \in {[}0,\pi)$ ($\psi \in {[}\pi,2\pi)$). 

\section{Pressure due to External Field}

\begin{figure}[htb]
    \centering
    \includegraphics[width=0.9\columnwidth]{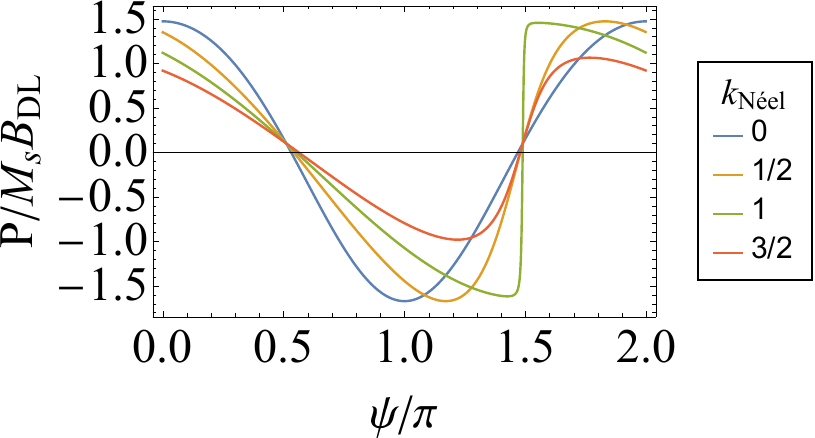}
    \caption{Normalized pressure $P/M_s\bdl$ for a N\'eel-like domain wall for several values of $k_{\text{N\'eel}}$, with $B_x = 10$ mT.}
    \label{fig:BxPvsPsiNeel}
\end{figure}

\begin{figure}[htb]
    \centering
    \includegraphics[width=0.9\columnwidth]{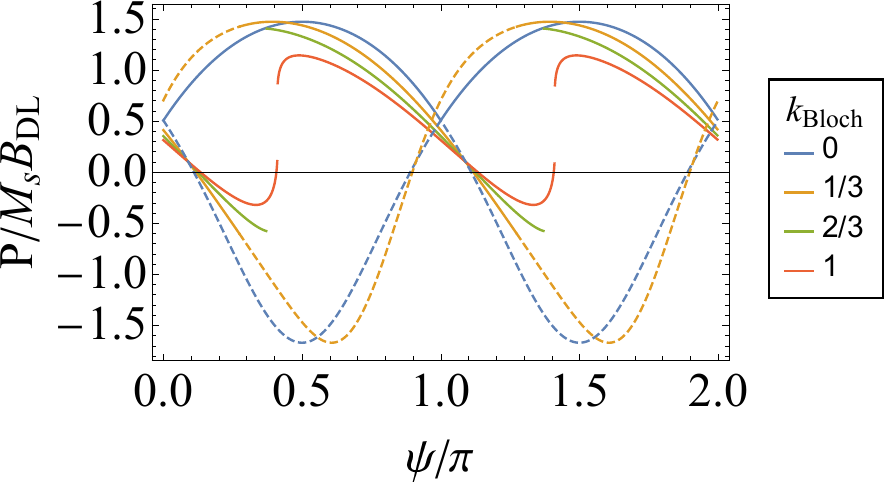}
    \caption{Normalized pressure $P/M_s\bdl$ for a Bloch-like domain wall for several values of $k_{\text{Bloch}}$, with $B_x = 10$ mT. Solid lines represent the lowest energy state; dashed lines the second lowest energy state.}
    \label{fig:BxPvsPsiBloch}
\end{figure}

\begin{figure}[htb]
    \centering
    \includegraphics[width=0.9\columnwidth]{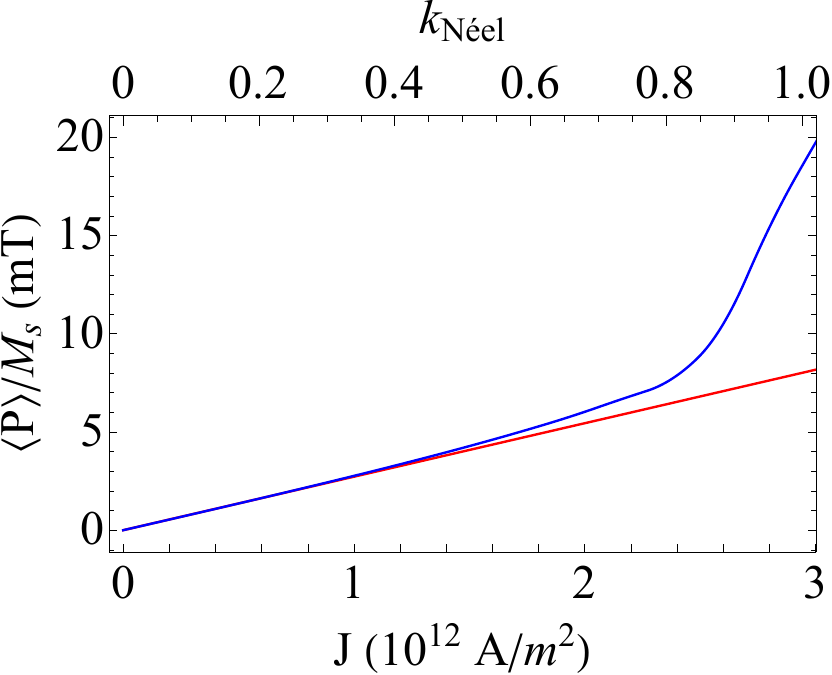}
    \caption{Pressure averaged over $\psi$ for the N\'eel-like domain wall and for material parameters: $M_s = 850\times 10^3$ A/m, $K = 1 $ MJ/m\textsuperscript{3}, $A_{ex} = 17$ pJ/m, $D = 1$ mJ/$\text{m}^2$, $\bdl=\bfl= 0.1$ T/($10^{11}$A/m\textsuperscript{2}), $B_x = 10$ mT. Blue is the exact numerical solution, red represents the perturbative solution for $k_{\text{N\'eel}} \ll 1$.}
    \label{fig:BxPavgNeel}
\end{figure}

\begin{figure}[htb]
    \centering
    \includegraphics[width=0.9\columnwidth]{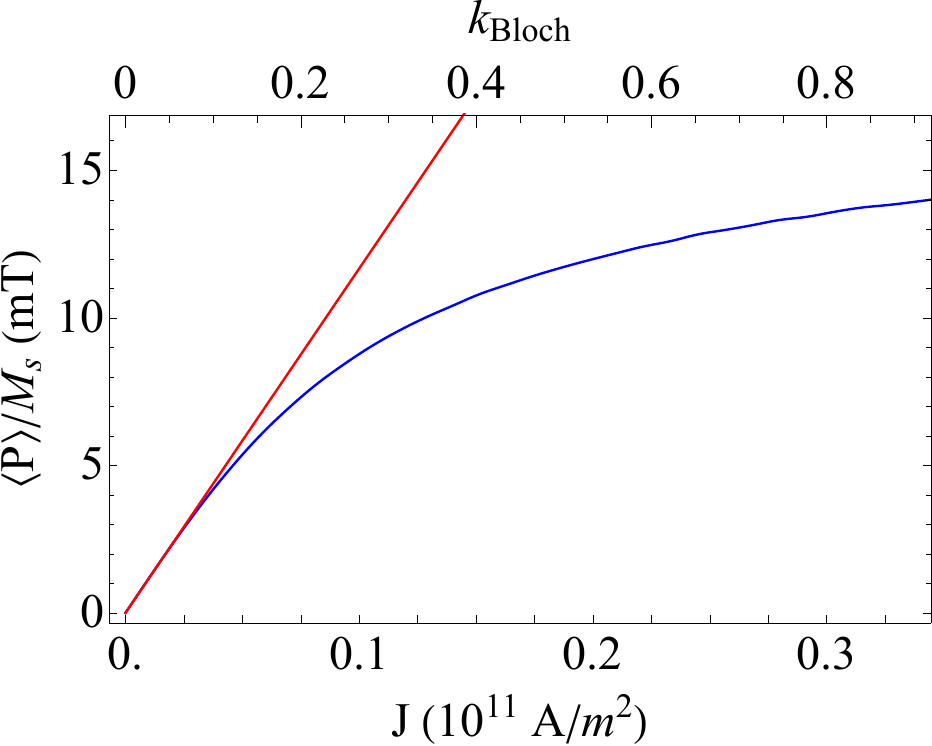}
    \caption{Pressure averaged over $\psi$ for the Bloch-like domain wall and for material parameters: $M_s = 850\times 10^3$ A/m, $K = 1 $ MJ/m\textsuperscript{3}, $A_{ex} = 17$ pJ/m, $t = 1$ nm, $\bdl=\bfl= 0.1$ T/($10^{11}$A/m\textsuperscript{2}), $B_x = 10$ mT. Blue is the exact numerical solution, red represents the perturbative solution for $k_{\text{Bloch}} \ll 1$.}
    \label{fig:BxPavgBloch}
\end{figure}

It is also useful to compare the pressure on domain walls due to 3m torques to the pressure without 3m torques in the presence of an applied magnetic field parallel to the direction of the current. Repeating our approach from sections \ref{sec:Neel} and \ref{sec:Bloch}, but now with $B_x \neq 0$ and $\bfltm = \bdltm = 0$ we arrive at transcendental equations for $\phi$;

\begin{equation}
    \frac{\sin\phi}{k_{\text{N\'eel}}\cos(\phi+\psi) + \frac{\Delta M_s B_x}{D}\sin(\psi+\phi)}=1
\end{equation}
in the N\'eel case and 
\begin{equation}
    \frac{\sin2\phi}{k_{\text{Bloch}}\cos(\psi+\phi)+\frac{\Delta \pi^2 B_x}{2\log2 \mu_0 M_s t}\sin(\psi+\phi)}=-1
\end{equation}
for the Bloch case.
We plot the normalized pressure as a function of $\psi$ in each case in Figs. \ref{fig:BxPvsPsiNeel} and \ref{fig:BxPvsPsiBloch} and the average pressure in Figs. \ref{fig:BxPavgNeel} and \ref{fig:BxPavgBloch}.

The most notable difference is that in both cases we see a linear dependence of the pressure applied current for small current densities. The reason for the linear (rather than quadratic dependence seen due to 3m torques) dependence at small currents can be easily understood; the external field which breaks the symmetry and allows for finite average pressure is independent of the applied current.

\section{Supporting Simulations}

The micromagnetic simulations in the main text were run with a relatively low temperature of 50K and large gilbert damping of $\alpha=0.5$ in order to accelerate convergence, reduce stochastic effects and improve the clarity of the discussion. In this section we discuss the impact of increased temperature and decreased gilbert damping on our findings.

Increasing the temperature of this system to 300K has several noteworthy effects, but does not lead to any qualitative difference in the outcomes of our simulations. The most important difference is that an increase of temperature increases the probability that the magnetic moment within domain walls fluctuates away from the energy minimum, which may impede the complete switching seen in the lower temperature case.

When the Gilbert damping is decreased from 0.5 to 0.05 the results for Bloch domain walls have no significant differences aside from a decrease in switching time. In the N\'eel case however there are some major qualitative differences.

For Gilbert damping of $\alpha = 0.5$ the perpendicular magnetization did not achieve saturation in greater than 95\% of cases, for small $k_{\text{N\'eel}}$. When $\alpha = 0.05$, perpendicular saturation was reached in most simulations. From this subset a small majority of final configurations resulted in a final value of $m_z=+1$, although this may be skewed due to the small number of disorder configurations. These results still agree with the discussion in section \ref{sec:Neel} which states that for small current densities $\bfltm$ does not lead to a positive pressure on domain walls, which prohibits deterministic magnetization reversal.

\begin{figure}[htb]
    \centering
    \includegraphics[width=0.9\columnwidth]{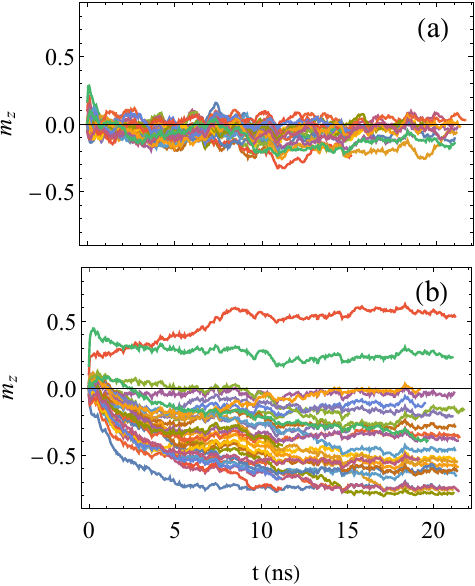}
    \caption{Micromagnetics results for $m_z$ vs time starting from 30 different random configurations $T= 300$ K, for (a) N\'eel-like domain walls for current density of $J = 2.5\times10^{10} $ A/m\textsuperscript{2} (b) Bloch-like domain walls for $J=2\times 10^{10}$ A/m\textsuperscript{2}. $M_s=\SI{850}{\kilo\ampere\per\meter}$, $A_{ex}=\SI{17}{\pico\joule\per\meter}$,  $K=\SI{0.5}{\mega\joule\per\cubic\meter}$, $\alpha=0.5$, $\bdl = \bfl = 0.1 $ T/($10^{11}$A/m\textsuperscript{2}), $\bfltm = 0.1 \bfl$.}
    \label{fig:mzvt_300K}
\end{figure}

\begin{figure}[htb]
    \centering
    \includegraphics[width=0.9\columnwidth]{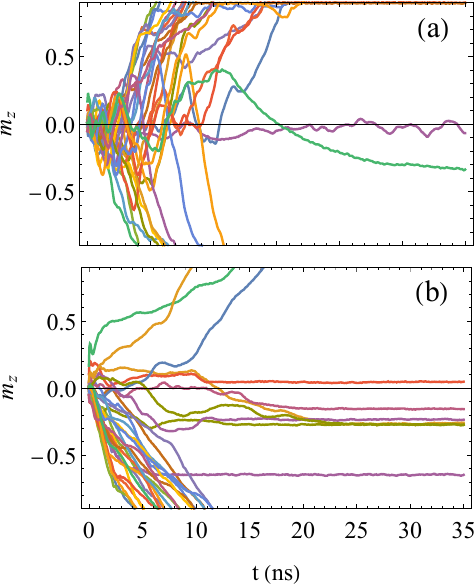}
    \caption{Micromagnetics results for $m_z$ vs time starting from 30 different random configurations with Gilbert damping $\alpha = 0.05$, for (a) N\'eel-like domain walls for current density of $J = 2.5\times10^{10} $ A/m\textsuperscript{2} (b) Bloch-like domain walls for $J=2\times 10^{10}$ A/m\textsuperscript{2}. $M_s=\SI{850}{\kilo\ampere\per\meter}$, $A_{ex}=\SI{17}{\pico\joule\per\meter}$,  $K=\SI{0.5}{\mega\joule\per\cubic\meter}$, $\bdl = \bfl = 0.1 $ T/($10^{11}$A/m\textsuperscript{2}), $\bfltm = 0.1 \bfl$, $T = 50$K.}
    \label{fig:mzvt_smallAlpha}
\end{figure}

\bibliography{Bib}

\end{document}